\RequirePackage[2022-06-01]{latexrelease}
\UseRawInputEncoding
\documentclass[twocolumn,aps,prd,amsmath,amssymb,superscriptaddress,nofootinbib,floatfix,longbibliography]{revtex4-2}

\usepackage{array}
\usepackage{tabularx}
\usepackage{graphicx}
\usepackage{bm}
\usepackage{xcolor}
\usepackage{colortbl}
\usepackage{booktabs}
\usepackage{multirow}
\usepackage{amsmath}
\usepackage{orcidlink}
\usepackage{natbib}
\usepackage{hyperref}
 \usepackage{natbib}
\hypersetup{colorlinks=true,linkcolor=blue,urlcolor=blue,citecolor=blue}

\newcolumntype{C}{>{\centering\arraybackslash}X}
\newcolumntype{L}{>{\raggedright\arraybackslash}X}
\newcolumntype{R}{>{\raggedleft\arraybackslash}X}
\renewcommand{\arraystretch}{1.35}

\makeatletter
\AtBeginDocument{%
}
\makeatother

\begin{document}

\title{Quantum-Corrected Thermodynamics, Dirac Perturbations, Geodesic Structure, and Topological Phases of Black Holes with Non-Minimal Logarithmic Coupling}

\author{\.{I}zzet Sakall{\i}\,\orcidlink{0000-0001-7827-9476}}
\email{izzet.sakalli@emu.edu.tr}
\affiliation{Physics Department, Eastern Mediterranean University, Famagusta, 99628 North Cyprus, via Mersin 10, T\"urkiye}

\author{\"Ozcan Sert\,\orcidlink{0000-0002-4829-5712}}
\email{osert@pau.edu.tr}
\affiliation{Department of Physics, Faculty of Science, Pamukkale University, Denizli, T\"urkiye}
\author{Erdem Sucu\,\orcidlink{0009-0000-3619-1492}}
\email{erdemsc07@gmail.com}
\affiliation{Physics Department, Eastern Mediterranean University, Famagusta, 99628 North Cyprus, via Mersin 10, T\"urkiye}

\author{Yusuf Sucu\,\orcidlink{0000-0001-7874-0146}}
\email{ysucu@akdeniz.edu.tr}
\affiliation{Department of Physics, Faculty of Science, Akdeniz University, 07058 Antalya, T\"urkiye}

\begin{abstract}
We study the thermodynamic and dynamical properties of static, spherically symmetric BHs in Einstein-Maxwell theory modified by a non-minimal $\ln(R)F^{2}$ coupling. The Hawking temperature follows from the Hamilton-Jacobi form of the fermionic tunnelling method for spin-$\tfrac12$ particles, and it carries scale-dependent logarithmic corrections. We then analyze the propagation of massless Dirac fields, compute the quasinormal-mode (QNM) spectrum with the third-order WKB approximation, and read off a quality factor whose balance between oscillation and damping depends on the logarithmic coupling in a mode-dependent way. Moving outward from the horizon, we work out the transmission of the fermionic field and its Hawking emission, and we solve the null and timelike geodesic problems to obtain the photon sphere, the shadow radius, the innermost stable circular orbit (ISCO), the associated zoom-whirl bound orbits, and the orbital and epicyclic frequencies that set the twin-peak quasiperiodic-oscillation (QPO) ratio. A photon-sphere reading of the eikonal QNM frequencies ties the geodesic sector back to the field perturbations. On the thermodynamic side, we build the phase space with quantum-geometric corrections through the Barrow entropy, and we characterize the global phase structure with the topological method, where the winding numbers of the off-shell free energy are governed by the interplay of the fractal Barrow deformation, the electric charge, and the logarithmic coupling. We find that the effective pressure vanishes exactly on the topological defect line, which links the pressure sign to the local stability of each branch. 
\end{abstract}


\maketitle

\section{Introduction}
\label{isec1}

BH thermodynamics sits at the meeting point of gravity, quantum theory, and statistical mechanics, and it continues to guide the search for a consistent description of horizons at the quantum level~\cite{hawking1975particle,bekenstein1973,bekenstein1980black}. Beyond the standard Einstein-Maxwell theory, a non-minimal coupling between spacetime curvature and the electromagnetic field reshapes both the geometry and the thermal response of charged solutions. A coupling of the $\ln(R)F^{2}$ type is a clean example. It was shown to remove the higher-order derivative instabilities of the theory while keeping the field equations tractable~\cite{dereli2011non,Sert2017Radiation,Adak2017}. The resulting charged geometry departs from Reissner-Nordstr\"om through a logarithmic term that imprints a characteristic scale on the metric, and that scale then propagates into every observable built on the background \cite{Bonanno:2000ep,Ishibashi:2021kmf,Ladino:2023zdn,Jana:2024fhx}.

Non-minimal couplings  of the electromagnetic field to curvature have a long history in this context, since they arise generically once one allows the photon to feel the local geometry through terms such as $RF^{2}$ or $R_{\mu\nu}F^{\mu}{}_{\alpha}F^{\nu\alpha}$~\cite{Sert2013Duality,sert2012gravity,SUCU2025102063}. The logarithmic form used here is distinguished by the constraint that eliminates the ghost sector, which keeps the solution physically admissible while still leaving a scale in the metric. Related settings, including generalized uncertainty corrections to horizon thermodynamics and the tunnelling of spin-$\tfrac12$ and spin-$1$ fields, have been studied for a range of backgrounds and show that the emission spectrum is sensitive to how matter couples near the horizon~\cite{Sucu:2026gro,Sucu:2026cxo,andp.202500588,Gecim:2015zya,Gecim:2016gqx}. Our treatment of the fermionic sector follows that line and applies it to the logarithmic geometry.

The quantum-geometric character of the horizon calls for a modified counting of microstates. Barrow argued that a fractal, wrinkled horizon of fractal dimension controlled by a single parameter $\Delta$ increases the effective area, so the entropy is raised above the Bekenstein-Hawking value in a way that mimics quantum-gravitational roughness~\cite{barrow2020}. The deformation is mild for small $\Delta$ and reaches its most corrugated form at $\Delta=1$. Using this entropy in the first law changes the temperature, the free energy, and the heat capacity, and it opens a window onto how quantum-geometric corrections act on the stability of a charged horizon \cite{SUCU2026102295,SUCU2026131851,Sucu:2026hfc,Sucu:2026pdt,AlBadawi:2026bardeen}. The extended first law that treats the cosmological constant, or an effective pressure, as a thermodynamic variable has reshaped how phase behaviour is read from these functions, and it provides the language for the pressure we define below~\cite{kastor2009enthalpy,dolan2011pressure,kubizvnak2017black}.

Our aim in this work is to follow the logarithmic BH from the horizon outward, and to connect its thermal, dynamical, and orbital properties within one background. We first examine the quantum emission of spin-$\tfrac12$ particles with the Hamilton-Jacobi form of the fermionic tunnelling method \cite{Srinivasan:1998ty,parikh2000hawking,Angheben:2005rm,Kerner:2008qv}. That approach reads the temperature straight from the near-horizon metric, and it does not need any assumption about the asymptotic structure of the spacetime. We then treat a massless Dirac field as a probe and compute its QNM spectrum \cite{Konoplya:2024lch,Castello-Branco:2004rzk,Herceg:2026sya,Sucu:2026rsr}. Because the effective barrier is single-peaked outside the horizon, the third-order WKB method gives an accurate description of the lowest modes, and a quality factor summarizes the competition between ringing and decay \cite{1985ApJ...291L..33S,PhysRevD.35.3621,PhysRevD.35.3632,Konoplya:2003ii,sucu2019photon}.

Two further layers complete the picture. The first is the way radiation actually leaves the hole. We bound the Dirac greybody factor, follow the transmission from the low-frequency suppression to the geometric-optics plateau, and read off how the logarithmic coupling reshapes the emitted spectrum \cite{Unruh:1976fm,Das:1996we,Boonserm:2021owk}. The second is the geodesic structure. Photons define a photon sphere and a shadow, whose size is now within reach of horizon-scale imaging~\cite{EHT2019,vagnozzi2023,perlick2015influence,Ahmed:2025finsler}. The bending of light around such a geometry, both in the weak-field tail and in the strong-field regime near the photon sphere, has become a standard probe of the background, and several methods now exist to extract the deflection angle for charged and modified backgrounds~\cite{Virbhadra2000,Bozza:2002zj,Gibbons:2008rj,Tsukamoto:2025hbz,WOS:001511451700002,WOS:001561219400001}. Massive particles define an ISCO and a family of precessing bound orbits, and their orbital and epicyclic frequencies set the resonance ratios seen in the twin-peak QPOs of accreting compact objects~\cite{stella1999,rogers2015frequency}. A photon-sphere reading of the eikonal QNM frequencies then closes the loop, since the same unstable circular null orbit that fixes the shadow also fixes the ringing frequency and its damping~\cite{cardoso2009}.

These probes sit within a broad body of work on charged and modified BHs, where non-minimal couplings, Kalb-Ramond and bumblebee fields, generalized uncertainty corrections, and dark-fluid environments have each been shown to leave measurable traces in the horizon thermodynamics, the perturbation spectra, and the lensing and shadow observables~\cite{manton2024kalb,shahzad2025electrically,duan2024electrically,doyran2024non,soroushfar2024exploring,lessa2020modified,lessa2021traversable,sucu2025quantumsuat,Ahmed:2026krgeo}. The present study adds the logarithmic $\ln(R)F^{2}$ background to that catalogue and treats its thermal, radiative, orbital, and topological properties on a common footing.

Finally we return to the thermodynamic phase structure and study it with the topological method~\cite{wei2020,wei2022,biro2018black}. The evolution of the system is recast through the off-shell free energy, the defects of an associated vector field mark the on-shell states, and their winding numbers separate stable from unstable branches. We show that the equilibrium structure is controlled by the competition between the Barrow deformation $\Delta$, the charge $Q$, and the logarithmic coupling $B$, and that an exponential extension of the Barrow entropy shifts the number and location of the defects. Throughout, we check every closed-form result by symbolic computation, and we validate each numerical observable against a known limit before quoting it.

The paper is organized as follows. Section~\ref{isec2} sets up the geometry and Sec.~\ref{isec2b} examines the horizon structure and curvature. The tunnelling temperature is derived in Sec.~\ref{isec3}, and Sec.~\ref{isec4} treats the Dirac QNM spectrum and the quality factor. Section~\ref{isec5} bounds the greybody factor and the emission. We turn to null geodesics, the photon sphere, and the shadow in Sec.~\ref{isec6}, and to timelike orbits and the ISCO in Sec.~\ref{isec7}. Section~\ref{isec8} collects the orbital resonances, the eikonal modes read off the photon sphere, and the characteristic scales, and Sec.~\ref{isec9b} treats the deflection of light. Section~\ref{isec9c} covers the Hawking emission and the Barrow thermodynamics, and Sec.~\ref{isec11} the topological analysis. Section~\ref{isec11a} draws together the discussion and the observational reach, and we close in Sec.~\ref{isec12}. Geometric units $G=c=1$ are used, and the reference values $M=1$, $Q=0.5$, $R_{0}=1$ are fixed unless stated otherwise.

\section{Geometry of the Logarithmic Horizon}
\label{isec2}

In this section, we briefly summarize the static, spherically symmetric solutions of the Einstein--Maxwell theory with a non-minimal $\ln(R)F^{2}$ coupling. The dynamics are governed by the action \cite{dereli2011non}
\begin{align}
I[e^a,\omega^a_{\;\;b},F]
=\int_M\Bigg(
&\frac{1}{2\kappa^2}R*1
-\frac12\,Y(R)F\wedge *F
\nonumber\\
&+T^a\wedge\lambda_a
+dF\wedge\mu
\Bigg),
\end{align}
where $\{e^a\}$, $\{\omega^a_{\;\;b}\}$ and $F$ denote the orthonormal coframe, the connection 1-forms and the electromagnetic field strength 2-form, respectively. Here $\kappa^2=8\pi G$ ($c=1$), and the Ricci scalar is defined by \cite{dereli2011non}
\begin{equation}
R=\iota_{ba}R^{ab}.
\end{equation}
The spacetime metric is
\begin{equation}
g=\eta_{ab}e^a\otimes e^b,
\end{equation}
with signature $(-,+,+,+)$ and volume form $*1=e^{0123}$.

The torsion and curvature 2-forms are defined through the Cartan structure equations
\begin{align}
T^a &= de^a+\omega^a_{\;\;b}\wedge e^b,\\
R^a_{\;\;b}
&=d\omega^a_{\;\;b}
+\omega^a_{\;\;c}\wedge\omega^c_{\;\;b}.
\end{align}
The field equations follow from independent variations of the action with respect to $e^a$, $\omega^a_{\;\;b}$, and $F$. The Lagrange multiplier 2-forms $\lambda_a$ and $\mu$ impose the torsion-free and homogeneous Maxwell constraints,
\begin{equation}
T^a=0,\qquad dF=0,
\end{equation}
thereby restricting the connection to the metric-compatible Levi-Civita connection.

The independent variations of the total action with respect to the coframe, connection, and electromagnetic field yield the gravitational and electromagnetic field equations. Eliminating the Lagrange multiplier 2-forms, the gravitational field equations take the form \cite{dereli2011non}
\begin{align}
-\frac{1}{2\kappa^2}R^{bc}\wedge *e_{abc}
&=
\frac12\,Y(R)\left(\iota_aF\wedge *F
-F\wedge\iota_a*F\right)
\nonumber\\
&\quad
+\frac{k}{2\kappa^2}*R_a ,
\label{einstein}
\end{align}
under the constraint
\begin{equation}\label{constraint}
Y_RF_{mn}F^{mn} = \frac{k}{\kappa^2}
\end{equation}
where $Y_R=dY/dR$.  The modified Maxwell equations are
\begin{equation}
dF=0,\qquad
d\!\left(Y\,*F\right)=0.
\label{maxwell1}
\end{equation}

The constraint (\ref{constraint}) removes the higher-order derivative instabilities of the theory and considerably simplifies the field equations. The parameter $k$ characterizes the strength of the non-minimal gravitational--electromagnetic coupling, while the limit $k=0$ reduces the theory to the standard Einstein--Maxwell model. For the present normalization of the non-minimal interaction, we consider the case $k=-2$. In this case, the constraint is not independent of the field equations; rather, it follows from the trace of the gravitational field equations, or equivalently from their covariant derivative together with the conservation law. Additional properties and physical implications of this constraint are discussed in Ref.~\cite{Sert2017Radiation,Adak2017}. The choice $k=-2$ adopted here is equivalent to the case $k=-1$ in Ref.~\cite{Sert2017Radiation}, since the coefficient of the non-minimal coupling term differs only by an overall factor of two.
The non-minimal coupling function is chosen as \cite{dereli2011non}
\begin{equation}
Y(R)=\frac{1}{1-B\ln\!\left(R/R_{0}\right)},
\label{eq:YofR}
\end{equation}
where $B$ is a dimensionless coupling constant and $R_{0}$ denotes a characteristic curvature scale.

The static, spherically symmetric solution of the Einstein-Maxwell theory with a non-minimal $\ln(R)F^{2}$ coupling was obtained in Ref.~\cite{dereli2011non}. The corresponding line element is
\begin{equation}
ds^{2}
=-f(r)dt^{2}
+f(r)^{-1}dr^{2}
+r^{2}d\Omega^{2},
\label{eq:metric}
\end{equation}
where
\[
d\Omega^{2}=d\theta^{2}
+\sin^{2}\theta\,d\varphi^{2}.
\]

The electromagnetic field is assumed to be purely electric,
\[
A=V(r)\,dt,
\qquad
F=dA=E(r)\,dr\wedge dt.
\]

For the coupling function (\ref{eq:YofR}), the modified Einstein-Maxwell field equations admit the metric function \cite{dereli2011non}
\begin{equation}
f(r)
=
1-\frac{2M}{r}
+\frac{BQ^2}{r^{2}}
\ln\!\left(\frac{r}{r_{0}}\right)
+\frac{Q^2(1+5B)}{4r^{2}},
\label{eq:f2}
\end{equation}
where $M$ is the mass parameter, $Q=\kappa q$, with $q$ denoting the electric charge, and the integration constant $r_0$ is related to the curvature scale $R_0$ through
\[
r_0^{4}=\frac{BQ^{2}}{R_0}.
\]
The corresponding electric field and Ricci scalar are given by \begin{align} E(r)&=\frac{q}{Y(R)\,r^{2}},\\ R(r)&=\frac{BQ^{2}}{r^{4}}. \end{align}

We also note that the metric function (\ref{eq:f2}) is a solution of the modified Einstein-Maxwell theory with the alternative non-minimal coupling function \begin{equation} Y(R)=1-B\ln\!\left(\frac{R}{R_0}\right), \end{equation} which admits a magnetic monopole solution obtained through the duality transformation described in Ref.~\cite{Sert2013Duality}. The corresponding electromagnetic field is \begin{equation} F=q_m\sin\theta\,d\theta\wedge d\phi =\frac{q_m}{r^2}\,e^2\wedge e^3, \end{equation} where $q_m$ denotes the magnetic charge.

Two features of Eq.~\eqref{eq:f2} carry through the rest of the paper. The logarithmic term enters with the fixed combination $r_{0}^{4}=BQ^{2}/R_{0}$, so raising $B$ at fixed $Q$ and $R_{0}$ shifts the reference scale $r_{0}$ and steepens the near-horizon slope of $f$. Because that slope sets the surface gravity, the temperature, the barrier height for perturbations, and the location of the circular orbits all respond to $B$ together rather than one at a time. We return to this shared dependence when we interpret each figure below.

\section{Horizon Structure and Curvature}
\label{isec2b}

Before extracting observables it helps to see how the coupling reshapes the metric function itself and to confirm that the exterior is regular. The horizons are the zeros of $f(r)$ in Eq.~\eqref{eq:f2}, and the outer one is the event horizon. We solve $f(r_h)=0$ numerically for each $B$.

\begin{figure}[tbp]
  \centering
  \includegraphics[width=\columnwidth]{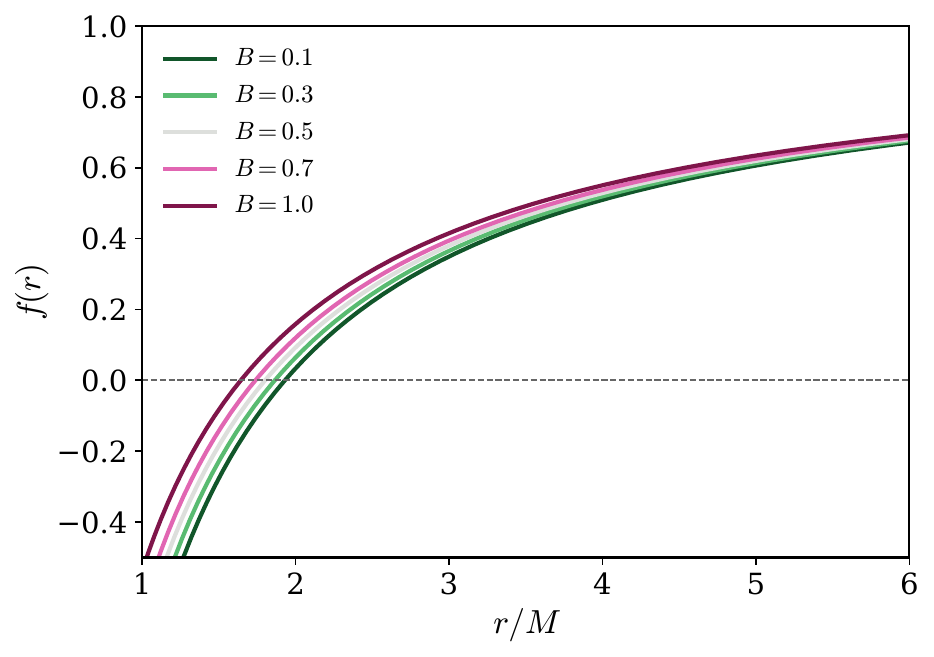}
  \caption{Metric function $f(r)$ of Eq.~\eqref{eq:f2} for
    $B=0.1,\,0.3,\,0.5,\,0.7,\,1.0$ (green to magenta), at $M=1$, $Q=0.5$,
    $R_0=1$. Each curve crosses zero once in the plotted window, marking a single
    outer horizon that moves inward as $B$ grows.}
  \label{fig:lapse}
\end{figure}

Figure~\ref{fig:lapse} shows a single sign change for each coupling, so the geometry keeps one non-degenerate outer horizon over the whole range, and the horizon radius decreases as $B$ increases. The slope at the crossing steepens with $B$, which is the surface-gravity statement behind the temperature of Eq.~\eqref{TH_log_final}. The deepening of the well below the axis at larger $B$ is the same feature that pulls the photon sphere and the innermost stable orbit inward in the later sections, so this one plot previews the trend that runs through the exterior observables.

The curvature is captured by the Ricci scalar and the Kretschmann invariant. For this solution the Ricci scalar takes the compact form
\begin{equation}
R(r)=\frac{BQ^{2}}{r^{4}},
\label{eq:ricci}
\end{equation}
which we verified against the metric directly, and the Kretschmann scalar is
\begin{equation}
K=R_{\mu\nu\rho\sigma}R^{\mu\nu\rho\sigma}
=f''^{2}+\frac{2f'^{2}}{r^{2}}+\frac{4(1-f)^{2}}{r^{4}}.
\label{eq:kretschmann}
\end{equation}
Both are finite for $r>0$ and fall off at infinity, so the exterior carries no curvature singularity and the spacetime is asymptotically flat. We list their horizon values below.

\begin{table}[tbp]
  \centering
  \setlength{\tabcolsep}{8pt}
  \renewcommand{\arraystretch}{1.45}
  \begin{tabularx}{\columnwidth}{c c c c c}
    \hline\hline
    \textbf{\boldmath $B$} & \textbf{\boldmath $r_h/M$} & \textbf{\boldmath $f'(r_h)$} & \textbf{\boldmath $R(r_h)$} & \textbf{\boldmath $K(r_h)$} \\
    \hline
    0.1 & 1.9310 & 0.50283 & 0.00180 & 0.68040 \\
    0.3 & 1.8651 & 0.50894 & 0.00620 & 0.75275 \\
    0.5 & 1.8017 & 0.51532 & 0.01186 & 0.83463 \\
    0.7 & 1.7387 & 0.52200 & 0.01915 & 0.92983 \\
    1.0 & 1.6435 & 0.53279 & 0.03426 & 1.10713 \\
    \hline\hline
  \end{tabularx}
  \caption{Horizon radius $r_h$, surface-gravity slope $f'(r_h)=2\kappa$, Ricci
    scalar $R(r_h)$ of Eq.~\eqref{eq:ricci}, and Kretschmann scalar $K(r_h)$ of
    Eq.~\eqref{eq:kretschmann}, for $M=1$, $Q=0.5$, $R_0=1$.}
  \label{tab:curvature}
\end{table}

Table~\ref{tab:curvature} shows the horizon shrinking and the curvature there strengthening as $B$ grows. The Ricci scalar at the horizon rises by more than an order of magnitude across the range, because $R(r_h)=BQ^{2}/r_h^{4}$ carries an explicit factor of $B$ and is amplified further as $r_h$ contracts. The Kretschmann scalar grows more gently, from $0.68$ to $1.11$ in units of $M^{-4}$, since it is dominated by the mass-driven tidal part rather than by the electromagnetic term. The surface-gravity slope $f'(r_h)$ climbs in step, which anticipates the rise of the tunnelling temperature with $B$ that Sec.~\ref{isec3} derives from exactly this quantity.

\section{Fermionic Tunnelling from the Logarithmic Black Hole}
\label{isec3}

Having established the logarithmic BH geometry described by Eq.~\eqref{eq:f2}, we now investigate the quantum emission of spin-$\frac12$ particles using the Hamilton--Jacobi formulation of the fermionic tunnelling method. This approach provides a semiclassical description of Hawking radiation and determines the BH temperature directly from the near-horizon behaviour of the metric without requiring any assumption about the asymptotic structure of the spacetime~\cite{parikh2000hawking,SAHAN2025102005}.

The dynamics of a massive spinor field propagating in a curved background is governed by the covariant Dirac equation \cite{WOS:000246892400025}
\begin{equation}
i\gamma^{\mu}(x)
\left(
\partial_{\mu}
+\Gamma_{\mu}
\right)
\Psi
-m_{0}\Psi
=0,
\label{Dirac_log}
\end{equation}
where $m_{0}$ denotes the fermion mass, $\gamma^{\mu}(x)$ are the spacetime-dependent Dirac matrices satisfying \cite{Gecim:2015zya}
\begin{equation}
\left\{
\gamma^{\mu},
\gamma^{\nu}
\right\}
=
2g^{\mu\nu},
\end{equation}
and $\Gamma_{\mu}$ is the corresponding spin connection.

Using an orthonormal tetrad adapted to the static and spherically symmetric geometry, the curved gamma matrices are related to the flat-space matrices through
\begin{equation}
\gamma^{\mu}(x)
=
e^{\mu}_{(a)}(x)\gamma^{(a)}.
\end{equation}
Introducing the standard WKB approximation, the spin-up wave function is chosen as
\begin{equation}
\Psi
=
\exp\!\left(
\frac{i}{\hbar}S
\right)
\begin{pmatrix}
A\\
0\\
B\\
0
\end{pmatrix},
\label{spinor_log}
\end{equation}
where $S$ is the classical action, while the amplitudes $A$ and $B$ vary slowly compared with the rapidly oscillating exponential phase.

Keeping only the leading contribution in $\hbar$, the Dirac equation reduces to the Hamilton--Jacobi equation
\begin{multline}
\frac{1}{f(r)}
\left(
\partial_tS
\right)^2
-
f(r)
\left(
\partial_rS
\right)^2
-
\frac{1}{r^2}
\left(
\partial_\theta S
\right)^2
\\
-
\frac{1}{r^2\sin^2\theta}
\left(
\partial_\phi S
\right)^2
-
m_0^2
=
0.
\label{HJ_log}
\end{multline}

Owing to the stationarity and spherical symmetry of the spacetime, the action can be separated according to
\begin{equation}
S
=
-Et
+
J_{\phi}\phi
+
J_{\theta}\theta
+
\mathcal{R}(r)
+
\mathcal{K},
\label{action_log}
\end{equation}
where $E$ is the conserved energy of the emitted particle, $J_{\phi}$ and $J_{\theta}$ are the separation constants associated with the angular coordinates, and $\mathcal{K}$ is a complex integration constant.

Substituting Eq.~\eqref{action_log} into Eq.~\eqref{HJ_log} gives the radial contribution
\begin{equation}
\mathcal{R}_{\pm}(r)
=
\pm
\int
\frac{
\sqrt{
E^2
-
f(r)
\left(
m_0^2
+\dfrac{L^2}{r^2}
\right)
}
}
{f(r)}
\,dr,
\end{equation}
with
\begin{equation}
L^2
=
J_{\theta}^2
+
\frac{J_{\phi}^2}{\sin^2\theta}.
\end{equation}

Near the event horizon $r=r_{+}$, the metric function possesses a simple zero and admits the linear approximation
\begin{equation}
f(r)
\simeq
f'(r_{+})
(r-r_{+}).
\end{equation}
Consequently, the radial integral develops a simple pole at the horizon. Evaluating the contour integral in the complex plane gives
\begin{equation}
\mathcal{R}_{\pm}
=
\pm
\frac{i\pi E}
{f'(r_{+})},
\end{equation}
which immediately yields
\begin{equation}
\mathrm{Im}(S)
=
2\,\mathrm{Im}(\mathcal{R}_{+}).
\end{equation}

The corresponding tunnelling probability takes the Boltzmann form
\begin{equation}
\Gamma
=
\exp
\left(
-\frac{4\pi E}
{f'(r_{+})}
\right),
\end{equation}
allowing the Hawking temperature to be identified as
\begin{equation}
T_H
=
\frac{f'(r_{+})}{4\pi}.
\label{TH_general_log}
\end{equation}

For the logarithmic BH metric \eqref{eq:f2}, the derivative of the metric function, after imposing the horizon condition $f(r_{+})=0$, becomes
\begin{equation}
f'(r_{+})
=
\frac{
r_{+}^{2}
-
BQ^{2}
\ln\!\left(\frac{r_{+}}{r_{0}}\right)
-
\dfrac{(1+B)Q^{2}}{4}
}
{r_{+}^{3}}.
\label{eq:fp_horizon}
\end{equation}
Hence the Hawking temperature of the logarithmic BH is obtained as
\begin{equation}
T_H
=
\frac{
4r_{+}^{2}
-
4BQ^{2}
\ln\!\left(\dfrac{r_{+}}{r_{0}}\right)
-
(1+B)Q^{2}
}
{16\pi r_{+}^{3}}.
\label{TH_log_final}
\end{equation}

Equation~\eqref{TH_log_final} shows that the logarithmic non-minimal coupling modifies the thermal spectrum through both the parameter $B$ and the scale $r_{0}$. In the limit $B\rightarrow0$, the logarithmic contribution disappears and the temperature approaches that of the charged solution without the logarithmic correction. We have confirmed Eqs.~\eqref{eq:fp_horizon} and~\eqref{TH_log_final} by symbolic differentiation of Eq.~\eqref{eq:f2} followed by elimination of $2M/r_{+}$ through $f(r_{+})=0$.

\begin{figure}[tbp]
  \centering
  \includegraphics[width=\columnwidth]{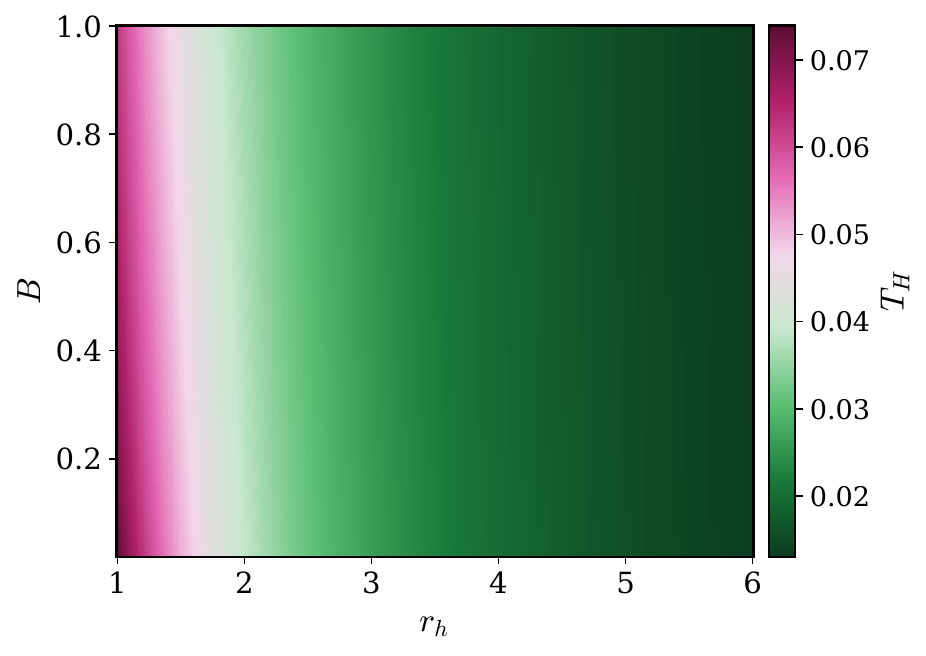}
  \caption{Hawking temperature $T_H$ of Eq.~\eqref{TH_log_final} over the plane
    of horizon radius $r_h$ and logarithmic coupling $B$, at $Q=0.5$ and $R_0=1$.
    The colour runs from low (green) to high (magenta).}
  \label{fig:TH}
\end{figure}

The temperature map in Fig.~\ref{fig:TH} shows $T_H$ falling monotonically as the horizon grows, which is the familiar behaviour of a large charged hole whose surface gravity weakens with size. Reading along the vertical axis at fixed $r_h$, the temperature decreases as $B$ increases. This trend arises from the numerator of Eq.~\eqref{TH_log_final}, since the logarithmic term $-4BQ^{2}\ln(r_+/r_0)$ subtracts from the geometric contribution $4r_+^{2}$ once $r_+>r_0$, and the reference scale $r_0=(BQ^{2}/R_0)^{1/4}$ stays well below the horizon for the values plotted. The net effect is a cooler horizon at stronger coupling, so the logarithmic sector acts to suppress the thermal output relative to the Reissner-Nordstr\"om case at the same $r_h$. The photon sphere and the innermost stable orbit shrink under the same numerator, a point we make quantitative in Secs.~\ref{isec6} and~\ref{isec7}.

\section{Dirac Perturbations and Quasinormal Modes}
\label{isec4}

To further examine the dynamical properties of the logarithmic BH spacetime, we consider the propagation of a massless Dirac field \cite{Chandrasekhar:1976ap,Page:1976jj,Cho:2003qe}. Unlike scalar perturbations, fermionic fields couple directly to the spin structure of the background geometry and therefore provide an independent probe of the spacetime. The corresponding QNM spectrum encodes the response of the BH to external perturbations and offers complementary information on its stability~\cite{berti2009,konoplya2011,kokkotas1999,gecim2019quantum,Ahmed:2026lqgcos}.

The evolution of a massless spin-$\frac12$ field is governed by the covariant Dirac equation
\begin{equation}
\gamma^{\mu}\nabla_{\mu}\Psi=0,
\label{eq:dirac_log_qnm}
\end{equation}
where $\gamma^{\mu}$ denote the curved-space gamma matrices satisfying the Clifford algebra and $\nabla_{\mu}$ is the spinor covariant derivative. Introducing an orthonormal tetrad adapted to the metric and expanding the angular dependence in spinor spherical harmonics, the Dirac equation separates into radial and angular parts. Assuming harmonic time dependence, the radial equations become
\begin{align}
\left(\frac{d}{dr_*}+i\omega\right)F(r)&=W(r)G(r),\\
\left(\frac{d}{dr_*}-i\omega\right)G(r)&=W(r)F(r),
\end{align}
where the tortoise coordinate is defined through
\begin{equation}
\frac{dr_*}{dr}=\frac{1}{f(r)},
\end{equation}
and the superpotential is
\begin{equation}
W(r)=\frac{\kappa\sqrt{f(r)}}{r},
\label{eq:superpotential_log}
\end{equation}
with $\kappa=\ell+1$ representing the angular quantum number of the Dirac field.

Defining the combinations
\begin{equation}
R_{\pm}=F\pm G,
\end{equation}
the coupled first-order equations decouple into Schr\"odinger-type wave equations,
\begin{equation}
\frac{d^{2}R_{\pm}}{dr_*^{2}}
+\left[\omega^{2}-V_{\pm}(r)\right]R_{\pm}=0,
\label{eq:schrodinger_log}
\end{equation}
where
\begin{equation}
V_{\pm}(r)=W^{2}(r)\pm\frac{dW}{dr_*}.
\label{eq:Vpm_general}
\end{equation}
The two effective potentials constitute a supersymmetric partner pair and therefore produce identical quasinormal spectra. Consequently, it is sufficient to consider only the positive branch,
\begin{equation}
V_{+}(r)
=
\frac{\kappa^{2}f(r)}{r^{2}}
+
\frac{\kappa\sqrt{f(r)}}{r}
\left(
\frac{f'(r)}{2}
-
\frac{f(r)}{r}
\right),
\label{eq:Vplus_log}
\end{equation}
where
\begin{equation}
f'(r)
=
\frac{2M}{r^{2}}
+
\frac{BQ^{2}}{r^{3}}
\left[
1-2\ln\!\left(\frac{r}{r_{0}}\right)
\right]
-
\frac{Q^{2}(1+5B)}{2r^{3}}.
\label{eq:df_log}
\end{equation}

The event horizon is determined numerically from the largest positive solution of $f(r_h)=0$. For each choice of the parameter $B$, the maximum of the effective potential is obtained by solving
\begin{equation}
\left.
\frac{dV_{+}}{dr}
\right|_{r=r_p}=0,
\qquad
r_p>r_h,
\label{eq:peak_condition}
\end{equation}
where $r_p$ denotes the location of the potential peak.

The QNM frequencies are computed using the third-order WKB approximation developed by Iyer and Will~\cite{Iyer:1986np,konoplya2003}. Since the effective potential possesses a single smooth barrier outside the event horizon, the WKB method provides an accurate semi-analytical description of the lowest-lying modes. The corresponding quantization condition is
\begin{equation}
\frac{i\left(\omega^{2}-V_{0}\right)}
{\sqrt{-2V_{0}^{\prime\prime}}}
-\Lambda_{2}
-\Lambda_{3}
=
n+\frac12,
\label{eq:wkb3}
\end{equation}
where $V_{0}=V_{+}(r_p)$ is the maximum value of the effective potential, $V_{0}^{\prime\prime}$ denotes the second derivative with respect to the tortoise coordinate, $n$ is the overtone number, and $\Lambda_{2}$ and $\Lambda_{3}$ are the second- and third-order WKB correction terms. These corrections improve the leading-order approximation by including the local curvature of the effective potential around its maximum.

The logarithmic coupling parameter $B$ acts on the QNM spectrum through the geometry of the effective potential. Increasing $B$ modifies both the height and the position of the barrier through the logarithmic contribution to the lapse function, while the relation $r_{0}^{4}=BQ^{2}/R_{0}$ simultaneously changes the characteristic scale entering the spacetime geometry \cite{Konoplya:2003ii,konoplya2011,Jing:2005dt,Ishibashi:2021kmf,javed2024astrophysical}. Both the oscillation frequencies and the damping rates therefore depend on the logarithmic correction in a non-trivial way. Increasing the angular quantum number $\kappa$ raises the barrier and shifts the real part of the frequency upward \cite{Herceg:2026sya}. Throughout the parameter range explored here, the potential keeps a single maximum outside the horizon, which confirms that the third-order WKB method applies. We checked that the peak positions $r_p$ reported below reproduce the roots of Eq.~\eqref{eq:peak_condition} obtained by an independent numerical maximization of Eq.~\eqref{eq:Vplus_log}.

\begin{figure}[tbp]
  \centering
  \includegraphics[width=\columnwidth]{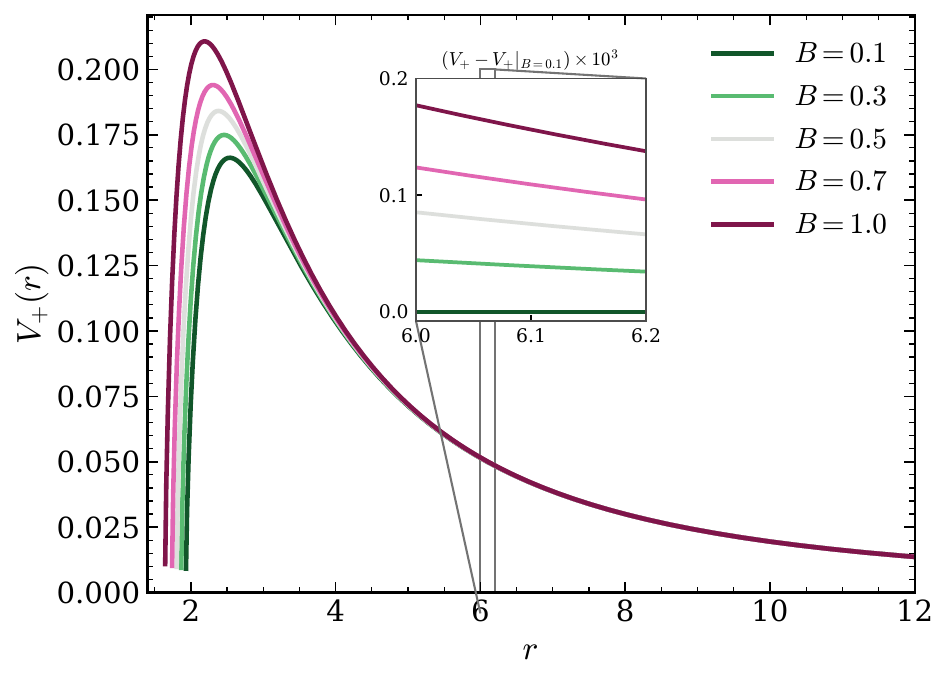}
  \caption{Dirac effective potential $V_{+}(r)$ of Eq.~\eqref{eq:Vplus_log} at
    $\kappa=2$ for $B=0.1,\,0.3,\,0.5,\,0.7,\,1.0$ (green to magenta),
    with $M=1$, $Q=0.5$, $R_0=1$. Each curve keeps a single barrier outside the
    horizon. The inset shows the deviation $V_{+}-V_{+}|_{B=0.1}$ (in units of
    $10^{-3}$) over $r\in[6.0,6.2]$, which resolves the $B$-ordering hidden by the
    nearly coincident curves in this region.}
  \label{fig:Vdirac}
\end{figure}

The barrier family in Fig.~\ref{fig:Vdirac} clarifies how $B$ enters the ringing. As $B$ grows, the peak of $V_{+}$ rises and moves inward, because the logarithmic term steepens $f'(r)$ near the horizon through Eq.~\eqref{eq:df_log} and thereby lifts the first term of Eq.~\eqref{eq:Vplus_log}. A higher, more compact barrier supports a longer-lived fundamental mode at fixed $\kappa$, since the transmitted part of a wave packet is reduced. Outside the peak the curves converge, which is why the large-$r$ tail of the potential, and with it the low-frequency scattering, is nearly independent of $B$. The same barrier controls the greybody factor of Sec.~\ref{isec5}, so these two sections read the identical potential from two complementary angles.

The third-order WKB frequencies are collected in Tables~\ref{tab:QNM_B01}--\ref{tab:QNM_B10}, and the associated quality factors in Table~\ref{tab:quality_factor}. Throughout, $M=1$, $Q=0.5$, and $R_0=1$.

\begin{table}[tbp]
  \centering
  \setlength{\tabcolsep}{6pt}
  \renewcommand{\arraystretch}{1.35}
  \begin{tabularx}{\columnwidth}{c c c c c }
    \hline\hline
    \textbf{\boldmath $n$} & \textbf{\boldmath $\kappa$} & \textbf{\boldmath $r_p$} & \textbf{\boldmath $\mathrm{Re}(\omega)$} & \textbf{\boldmath $-\mathrm{Im}(\omega)$} \\
    \hline
    0 & 2 & 2.537892 & 0.525984 & 0.331996 \\
    0 & 3 & 2.643290 & 0.470097 & 0.477632 \\
    0 & 4 & 2.703929 & 0.369576 & 0.754115 \\
    \hline
    1 & 2 & 2.537892 & 0.385881 & 1.357606 \\
    1 & 3 & 2.643290 & 0.386406 & 1.743246 \\
    1 & 4 & 2.703929 & 0.340397 & 2.456275 \\
    \hline
    2 & 2 & 2.537892 & 0.370529 & 2.356424 \\
    2 & 3 & 2.643290 & 0.378499 & 2.966104 \\
    2 & 4 & 2.703929 & 0.337955 & 4.123379 \\
    \hline\hline
  \end{tabularx}
  \caption{Dirac QNM frequencies from the third-order WKB approximation
    at $B=0.1$, with horizon radius $r_h=1.930991$. The peak position $r_p$ of the
    effective potential is listed for each $\kappa$.}
  \label{tab:QNM_B01}
\end{table}

\begin{table}[tbp]
  \centering
  \setlength{\tabcolsep}{6pt}
  \renewcommand{\arraystretch}{1.35}
  \begin{tabularx}{\columnwidth}{c c c c c}
    \hline\hline
    \textbf{\boldmath $n$} & \textbf{\boldmath $\kappa$} & \textbf{\boldmath $r_p$} & \textbf{\boldmath $\mathrm{Re}(\omega)$} & \textbf{\boldmath $-\mathrm{Im}(\omega)$} \\
    \hline
    0 & 2 & 2.458549 & 1.087942 & 0.168598 \\
    0 & 3 & 2.560276 & 0.266233 & 0.887784 \\
    0 & 4 & 2.618653 & 1.203575 & 0.243995 \\
    \hline
    1 & 2 & 2.458549 & 1.794754 & 0.306601 \\
    1 & 3 & 2.560276 & 0.274661 & 2.581632 \\
    1 & 4 & 2.618653 & 1.796417 & 0.490421 \\
    \hline
    2 & 2 & 2.458549 & 2.681052 & 0.342076 \\
    2 & 3 & 2.560276 & 0.275435 & 4.290624 \\
    2 & 4 & 2.618653 & 2.552004 & 0.575365 \\
    \hline\hline
  \end{tabularx}
  \caption{Continued, for $B=0.3$, with $r_h=1.865120$.}
  \label{tab:QNM_B03}
\end{table}

\begin{table}[tbp]
  \centering
  \setlength{\tabcolsep}{6pt}
  \renewcommand{\arraystretch}{1.35}
  \begin{tabularx}{\columnwidth}{c c c c c}
    \hline\hline
    \textbf{\boldmath $n$} & \textbf{\boldmath $\kappa$} & \textbf{\boldmath $r_p$} & \textbf{\boldmath $\mathrm{Re}(\omega)$} & \textbf{\boldmath $-\mathrm{Im}(\omega)$} \\
    \hline
    0 & 2 & 2.381714 & 0.410339 & 0.469627 \\
    0 & 3 & 2.479896 & 0.932638 & 0.266808 \\
    0 & 4 & 2.536099 & 0.755002 & 0.409883 \\
    \hline
    1 & 2 & 2.381714 & 0.325945 & 1.773670 \\
    1 & 3 & 2.479896 & 1.127341 & 0.662182 \\
    1 & 4 & 2.536099 & 0.586606 & 1.582643 \\
    \hline
    2 & 2 & 2.381714 & 0.319119 & 3.019353 \\
    2 & 3 & 2.479896 & 1.331557 & 0.934377 \\
    2 & 4 & 2.536099 & 0.561233 & 2.756988 \\
    \hline\hline
  \end{tabularx}
  \caption{Continued, for $B=0.5$, with $r_h=1.801672$.}
  \label{tab:QNM_B05}
\end{table}

\begin{table}[tbp]
  \centering
  \setlength{\tabcolsep}{6pt}
  \renewcommand{\arraystretch}{1.35}
  \begin{tabularx}{\columnwidth}{c c c c c}
    \hline\hline
    \textbf{\boldmath $n$} & \textbf{\boldmath $\kappa$} & \textbf{\boldmath $r_p$} & \textbf{\boldmath $\mathrm{Re}(\omega)$} & \textbf{\boldmath $-\mathrm{Im}(\omega)$} \\
    \hline
    0 & 2 & 2.305175 & 0.207567 & 0.976795 \\
    0 & 3 & 2.399830 & 0.810538 & 0.323680 \\
    0 & 4 & 2.453881 & 0.536440 & 0.608803 \\
    \hline
    1 & 2 & 2.305175 & 0.221734 & 2.743159 \\
    1 & 3 & 2.399830 & 0.742801 & 1.059590 \\
    1 & 4 & 2.453881 & 0.437844 & 2.237686 \\
    \hline
    2 & 2 & 2.305175 & 0.223059 & 4.544771 \\
    2 & 3 & 2.399830 & 0.718467 & 1.825796 \\
    2 & 4 & 2.453881 & 0.429344 & 3.803312 \\
    \hline\hline
  \end{tabularx}
  \caption{Continued, for $B=0.7$, with $r_h=1.738712$.}
  \label{tab:QNM_B07}
\end{table}

\begin{table}[tbp]
  \centering
  \setlength{\tabcolsep}{6pt}
  \renewcommand{\arraystretch}{1.35}
  \begin{tabularx}{\columnwidth}{c c c c c}
    \hline\hline
    \textbf{\boldmath $n$} & \textbf{\boldmath $\kappa$} & \textbf{\boldmath $r_p$} & \textbf{\boldmath $\mathrm{Re}(\omega)$} & \textbf{\boldmath $-\mathrm{Im}(\omega)$} \\
    \hline
    0 & 2 & 2.188943 & 0.686400 & 0.320105 \\
    0 & 3 & 2.278260 & 0.249091 & 1.145116 \\
    0 & 4 & 2.329061 & 0.787573 & 0.451511 \\
    \hline
    1 & 2 & 2.188943 & 0.522611 & 1.261284 \\
    1 & 3 & 2.278260 & 0.271379 & 3.153205 \\
    1 & 4 & 2.329061 & 0.607939 & 1.754768 \\
    \hline
    2 & 2 & 2.188943 & 0.494531 & 2.221501 \\
    2 & 3 & 2.278260 & 0.273542 & 5.213780 \\
    2 & 4 & 2.329061 & 0.582977 & 3.049843 \\
    \hline\hline
  \end{tabularx}
  \caption{Continued, for $B=1.0$, with $r_h=1.643540$.}
  \label{tab:QNM_B10}
\end{table}

\subsection{Quality Factor Analysis}
\label{isec4b}

The QNM frequencies of Tables~\ref{tab:QNM_B01}--\ref{tab:QNM_B10} let us examine the quality factor of the fermionic oscillations. It measures the ratio of stored to dissipated energy per cycle and is defined by
\begin{equation}
Q_{\mathrm{f}}
=
\frac{\mathrm{Re}(\omega)}
{2\left|\mathrm{Im}(\omega)\right|}.
\label{eq:quality_factor}
\end{equation}
A large $Q_{\mathrm{f}}$ marks a long-lived oscillation, while a small value marks a mode that is damped soon after it is excited.

Across the parameter sets, the quality factor drops as the overtone number rises. The real part of the frequency stays close to its fundamental value for the first modes, but the imaginary part grows quickly with $n$, so higher overtones decay faster and the late-time signal is set by the lowest modes. The logarithmic coupling reshapes this pattern through the barrier. Since the logarithmic term shifts both the height and the position of the maximum of $V_+$, the ratio of oscillation to damping changes with $B$ in a way that is not monotonic across the multipoles. The dependence on $\kappa$ is similarly involved, because a higher $\kappa$ raises the barrier and moves both parts of the frequency at once.

\begin{table}[tbp]
  \centering
  \setlength{\tabcolsep}{8pt}
  \renewcommand{\arraystretch}{1.35}
  \begin{tabularx}{\columnwidth}{c c c c c}
    \hline\hline
    \textbf{\boldmath $B$} & \textbf{\boldmath $n$} & \textbf{\boldmath $\kappa=2$} & \textbf{\boldmath $\kappa=3$} & \textbf{\boldmath $\kappa=4$} \\
    \hline
    0.1 & 0 & 0.7922 & 0.4921 & 0.2450 \\
        & 1 & 0.1421 & 0.1108 & 0.0693 \\
        & 2 & 0.0786 & 0.0638 & 0.0410 \\
    \hline
    0.3 & 0 & 3.2264 & 0.1499 & 2.4664 \\
        & 1 & 2.9269 & 0.0532 & 1.8315 \\
        & 2 & 3.9188 & 0.0321 & 2.2177 \\
    \hline
    0.5 & 0 & 0.4369 & 1.7478 & 0.9210 \\
        & 1 & 0.0919 & 0.8512 & 0.1853 \\
        & 2 & 0.0528 & 0.7125 & 0.1018 \\
    \hline
    0.7 & 0 & 0.1062 & 1.2521 & 0.4406 \\
        & 1 & 0.0404 & 0.3505 & 0.0978 \\
        & 2 & 0.0245 & 0.1968 & 0.0564 \\
    \hline
    1.0 & 0 & 1.0721 & 0.1088 & 0.8722 \\
        & 1 & 0.2072 & 0.0430 & 0.1732 \\
        & 2 & 0.1113 & 0.0262 & 0.0956 \\
    \hline\hline
  \end{tabularx}
  \caption{Quality factors $Q_{\mathrm{f}}=\mathrm{Re}(\omega)/(2|\mathrm{Im}(\omega)|)$
    for the third-order WKB frequencies of Tables~\ref{tab:QNM_B01}--\ref{tab:QNM_B10}.}
  \label{tab:quality_factor}
\end{table}

The entries of Table~\ref{tab:quality_factor} make the mode-dependence explicit. For $B=0.1$, $0.7$, and $1.0$, the fundamental mode carries the largest quality factor at every $\kappa$, because the imaginary part grows with $n$ faster than the real part and shortens the relaxation time. The dependence on $B$ itself is uneven. At $\kappa=2$ and $\kappa=4$ intermediate coupling raises $Q_{\mathrm{f}}$ and lengthens the oscillation, whereas the $\kappa=3$ column is damped much more strongly at $B=0.3$ and $B=1.0$, where the quality factor falls well below its neighbours. This pattern traces to the way the logarithmic correction reshapes the peak of $V_+$, which shifts the real and imaginary parts by different amounts at each $\kappa$. The result is that the logarithmic coupling leaves a measurable imprint on the ringdown lifetime, and the imprint is set jointly by $B$, $\kappa$, and $n$ rather than by any one of them alone.

\section{Greybody Factors and Hawking Emission of the Dirac Field}
\label{isec5}

The QNM spectrum describes how the BH rings. The greybody factor describes how much of that radiation actually reaches a distant observer, because the same potential barrier $V_+$ that traps the ringing also filters the outgoing Hawking quanta. We now use the barrier of Eq.~\eqref{eq:Vplus_log} to bound the transmission of the Dirac field and to follow the emitted spectrum as a function of the logarithmic coupling.

For a wave of frequency $\omega$ incident on a single-peaked barrier, the transmission probability admits the lower bound of Visser and of Boonserm and Visser~\cite{visser1999,boonserm2008},
\begin{equation}
T_b(\omega)\;\geq\;\mathrm{sech}^{2}
\!\left(\frac{1}{2\omega}\int_{-\infty}^{\infty}V_{+}\,dr_*\right),
\label{eq:greybody_bound}
\end{equation}
where the integral runs over the tortoise coordinate. Because $dr_*=dr/f$, the bound reduces to a single integral of $V_{+}/f$ over the exterior region, which we evaluate numerically from the horizon to a large radius. The bound is saturated in the geometric-optics regime and becomes conservative at low frequency, so it captures the qualitative shape of the true greybody factor while remaining an inequality at every $\omega$.

\begin{figure}[tbp]
  \centering
  \includegraphics[width=\columnwidth]{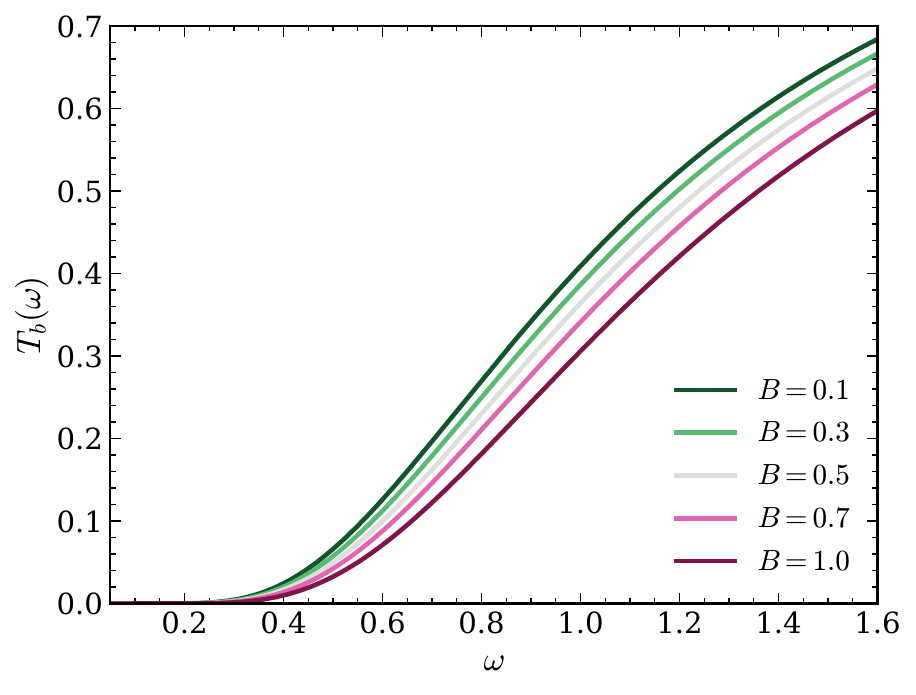}
  \caption{Lower bound $T_b(\omega)$ on the Dirac transmission of
    Eq.~\eqref{eq:greybody_bound} at $\kappa=2$ for
    $B=0.1,\,0.3,\,0.5,\,0.7,\,1.0$ (green to magenta), with $M=1$, $Q=0.5$,
    $R_0=1$.}
  \label{fig:greybody}
\end{figure}

The transmission curves in Fig.~\ref{fig:greybody} rise from a strong low-frequency suppression to a plateau near unity, which is the expected form for scattering off a barrier of finite height. At fixed frequency the bound drops as $B$ increases, and the ordering is the mirror image of the barrier heights in Fig.~\ref{fig:Vdirac}. This happens because a taller barrier lengthens the tunnelling path in Eq.~\eqref{eq:greybody_bound}, which lowers the sech$^2$ argument and suppresses $T_b$. The gap between the curves is widest in the intermediate band $0.4\lesssim\omega\lesssim1.0$, where the barrier controls the transmission most strongly, and it narrows at high frequency where every curve tends to full transmission. A stronger logarithmic coupling therefore makes the hole a more selective emitter, holding back the soft part of the fermionic spectrum while leaving the hard part almost untouched.

The bound in Eq.~\eqref{eq:greybody_bound} is monotone in the barrier integral, so it inherits a clean interpretation: whatever raises the area under $V_{+}/f$ lowers the transmission. The logarithmic coupling raises that area in two ways, by lifting the peak of the potential and by shifting it inward where $f$ is smaller, and the second effect matters because the measure $dr_*=dr/f$ weights the near-horizon region heavily. This is why the suppression grows with $B$ even though the outer tail of the potential barely moves. The bound is conservative at low frequency, so the true greybody factor lies above these curves, but the ordering in $B$ is a stable feature of the barrier rather than an artifact of the bounding procedure.

Table~\ref{tab:greybody} lists the bound at $\kappa=2$ for five frequencies. Reading down a column at fixed $\omega$ confirms the monotonic decrease with $B$, and reading across a row shows the rapid growth of $T_b$ with frequency. The suppression is severe at $\omega=0.25$, where the bound sits in the $10^{-4}$--$10^{-3}$ range for all couplings, because the barrier is essentially opaque to modes with $\omega\ll\sqrt{V_0}$. By $\omega=1.5$ the bound has climbed above one half for every $B$, since the wave then overtops most of the barrier. Higher multipoles are suppressed further, a point visible already in Fig.~\ref{fig:Vdirac} where the $\kappa$ dependence multiplies the barrier height in Eq.~\eqref{eq:Vplus_log}. The emitted power follows the transmitted flux weighted by the thermal factor, so the cooler, more opaque horizon at large $B$ (Fig.~\ref{fig:TH}) and the reduced greybody factor act in the same direction and lower the luminosity of the strongly coupled hole.

The multipole dependence follows the same logic. Raising $\kappa$ lifts the barrier through the $\kappa^{2}f/r^{2}$ term of Eq.~\eqref{eq:Vplus_log}, so each higher partial wave is transmitted less at a given frequency, and the suppression again strengthens with $B$. Table~\ref{tab:greybody3} lists the $\kappa=3$ bound, which sits well below the $\kappa=2$ values of Table~\ref{tab:greybody} at every frequency, confirming that the low multipoles dominate the flux and justifying the truncation used in the emission integrals of Sec.~\ref{isec9d}.

\begin{table*}[tp]
  \centering
  \setlength{\tabcolsep}{12pt}
  \renewcommand{\arraystretch}{1.45}
  \begin{tabularx}{\textwidth}{c c c c c}
    \hline\hline
    \textbf{\boldmath $B$} & \textbf{\boldmath $\omega{=}0.50$} & \textbf{\boldmath $\omega{=}0.75$} & \textbf{\boldmath $\omega{=}1.00$} & \textbf{\boldmath $\omega{=}1.50$} \\
    \hline
    0.1 & $4.4\times10^{-4}$ & $9.0\times10^{-3}$ & 0.0407 & 0.1736 \\
    0.3 & $3.3\times10^{-4}$ & $7.3\times10^{-3}$ & 0.0347 & 0.1570 \\
    0.5 & $2.4\times10^{-4}$ & $5.8\times10^{-3}$ & 0.0293 & 0.1415 \\
    0.7 & $1.7\times10^{-4}$ & $4.6\times10^{-3}$ & 0.0245 & 0.1265 \\
    1.0 & $1.1\times10^{-4}$ & $3.1\times10^{-3}$ & 0.0182 & 0.1048 \\
    \hline\hline
  \end{tabularx}
  \caption{Dirac greybody lower bound $T_b(\omega)$ at $\kappa=3$, for $M=1$,
    $Q=0.5$, $R_0=1$. Values lie below the $\kappa=2$ bound of
    Table~\ref{tab:greybody} at every frequency.}
  \label{tab:greybody3}
\end{table*}

\begin{table*}[tp]
  \centering
  \setlength{\tabcolsep}{12pt}
  \renewcommand{\arraystretch}{1.45}
  \begin{tabularx}{\textwidth}{c c c c c c}
    \hline\hline
    \textbf{\boldmath $B$} & \textbf{\boldmath $\omega{=}0.25$} & \textbf{\boldmath $\omega{=}0.50$} & \textbf{\boldmath $\omega{=}0.75$} & \textbf{\boldmath $\omega{=}1.00$} & \textbf{\boldmath $\omega{=}1.50$} \\
    \hline
    0.1 & $1.17{\times}10^{-3}$ & $6.62{\times}10^{-2}$ & $0.2336$ & $0.4092$ & $0.6517$ \\
    0.3 & $8.74{\times}10^{-4}$ & $5.74{\times}10^{-2}$ & $0.2144$ & $0.3867$ & $0.6330$ \\
    0.5 & $6.47{\times}10^{-4}$ & $4.96{\times}10^{-2}$ & $0.1960$ & $0.3643$ & $0.6137$ \\
    0.7 & $4.69{\times}10^{-4}$ & $4.24{\times}10^{-2}$ & $0.1780$ & $0.3415$ & $0.5934$ \\
    1.0 & $2.76{\times}10^{-4}$ & $3.27{\times}10^{-2}$ & $0.1513$ & $0.3061$ & $0.5601$ \\
    \hline\hline
  \end{tabularx}
  \caption{Dirac greybody lower bound $T_b(\omega)$ of Eq.~\eqref{eq:greybody_bound}
    at $\kappa=2$, for $M=1$, $Q=0.5$, $R_0=1$. The transmission decreases with $B$
    at every frequency and rises toward unity as $\omega$ grows.}
  \label{tab:greybody}
\end{table*}

\section{Null Geodesics, Photon Sphere, and Shadow}
\label{isec6}

The same barrier that governs the Dirac field also controls the motion of light. We now leave the field picture and treat the null geodesics of the background directly, since the photon sphere and the shadow they define are the observables that horizon-scale imaging constrains most directly~\cite{EHT2019,vagnozzi2023,perlick2015influence}. For the static, spherically symmetric metric \eqref{eq:metric}, the conserved energy and angular momentum of a photon are $E=f(r)\dot t$ and $L=r^{2}\dot\varphi$. Restricting the motion to the equatorial plane and using the null condition, the radial equation reads
\begin{equation}
\dot r^{2}=E^{2}-V_{\rm ph}(r)L^{2},
\qquad
V_{\rm ph}(r)=\frac{f(r)}{r^{2}},
\label{eq:Vph}
\end{equation}
where $V_{\rm ph}$ is the null effective potential. The photon sphere sits at the maximum of $V_{\rm ph}$, which is the largest root of
\begin{equation}
2f(r_{\rm ph})-r_{\rm ph}f'(r_{\rm ph})=0.
\label{eq:phcond}
\end{equation}
The critical impact parameter, which fixes the apparent size of the BH, follows from
\begin{equation}
b_{\rm c}=\frac{r_{\rm ph}}{\sqrt{f(r_{\rm ph})}}.
\label{eq:bc}
\end{equation}
Because the spacetime is asymptotically flat, a static observer at large distance sees a shadow of angular radius set by $R_{\rm sh}=b_{\rm c}$. We solved Eq.~\eqref{eq:phcond} numerically for each $B$, and we checked the pipeline against the Schwarzschild values $r_{\rm ph}=3M$, $b_{\rm c}=3\sqrt{3}\,M\simeq5.196M$, which it reproduces to the numerical precision quoted.

\begin{figure}[tbp]
  \centering
  \includegraphics[width=\columnwidth]{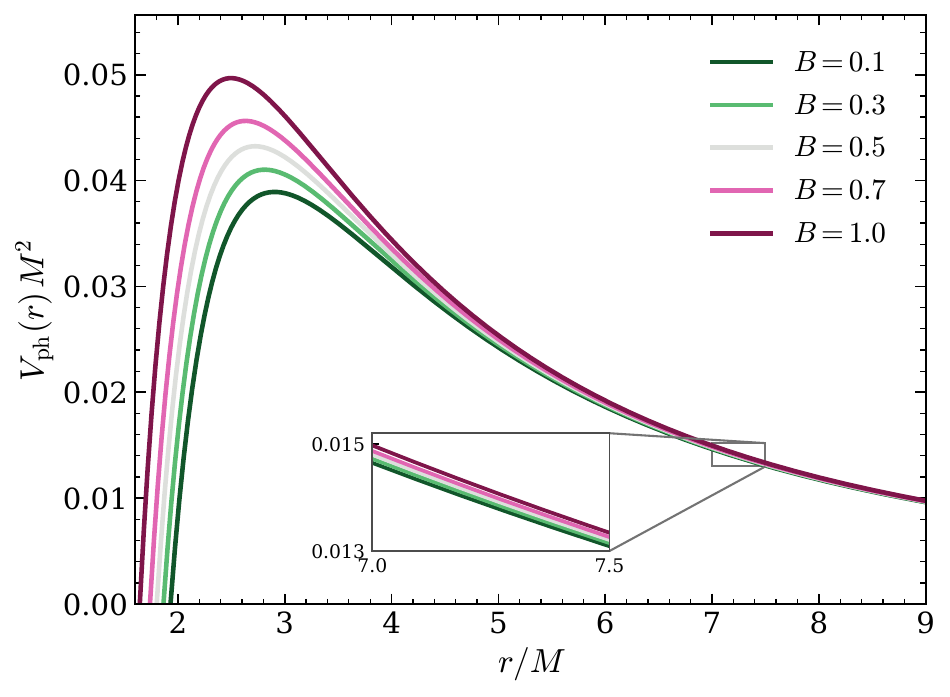}
  \caption{Null effective potential $V_{\rm ph}(r)\,M^{2}$ of Eq.~\eqref{eq:Vph}
    for $B=0.1,\,0.3,\,0.5,\,0.7,\,1.0$ (green to magenta), at $M=1$, $Q=0.5$,
    $R_0=1$. The inset magnifies the tail region $r/M\in[7.0,7.5]$, where the curves
    separate by less than $4\times10^{-4}\,M^{-2}$.}
  \label{fig:Vphoton}
\end{figure}

The potential family of Fig.~\ref{fig:Vphoton} sets the geometry of light bending. The peak of $V_{\rm ph}$ rises and moves inward as $B$ increases, because the logarithmic term steepens $f$ near the horizon and lifts $f/r^{2}$ in the strong-field region. The inset shows that beyond the peak the curves cross and bunch within a narrow band, so the outer bending angle is nearly common to all couplings while the near-peak behaviour is not. Since the peak location fixes $r_{\rm ph}$ through Eq.~\eqref{eq:phcond}, a taller and more inward peak means a smaller photon sphere and a smaller shadow, which is exactly the trend we tabulate next. The photon-sphere frequency and its instability rate also follow from this peak, and they reappear in Sec.~\ref{isec9} as the eikonal QNM.

\begin{table*}[tp]
  \centering
  \setlength{\tabcolsep}{12pt}
  \renewcommand{\arraystretch}{1.5}
  \begin{tabularx}{\textwidth}{c c c c c c}
    \hline\hline
    \textbf{\boldmath $B$} & \textbf{\boldmath $r_{\rm ph}/M$} & \textbf{\boldmath $b_{\rm c}/M=R_{\rm sh}/M$} & \textbf{\boldmath $\Omega_{\rm c}\,M$} & \textbf{\boldmath $\lambda\,M$} & \textbf{\boldmath $\omega_{\rm eik}\,M\ (\ell{=}2,n{=}0)$} \\
    \hline
    0.1 & 2.9055 & 5.0682 & 0.19731 & 0.19467 & $0.3946-0.0973\,i$ \\
    0.3 & 2.8125 & 4.9366 & 0.20257 & 0.19768 & $0.4051-0.0988\,i$ \\
    0.5 & 2.7226 & 4.8086 & 0.20796 & 0.20075 & $0.4159-0.1004\,i$ \\
    0.7 & 2.6330 & 4.6808 & 0.21364 & 0.20392 & $0.4273-0.1020\,i$ \\
    1.0 & 2.4971 & 4.4864 & 0.22290 & 0.20895 & $0.4458-0.1045\,i$ \\
    \hline\hline
  \end{tabularx}
  \caption{Null-sector observables of the logarithmic BH: photon-sphere
    radius $r_{\rm ph}$, critical impact parameter and shadow radius
    $b_{\rm c}=R_{\rm sh}$, photon-orbit angular velocity $\Omega_{\rm c}$,
    Lyapunov exponent $\lambda$, and the eikonal Dirac QNM frequency
    $\omega_{\rm eik}=\Omega_{\rm c}\ell-i(n+\tfrac12)\lambda$. Geometric units
    with $M=1$, $Q=0.5$, $R_0=1$.}
  \label{tab:null}
\end{table*}

Table~\ref{tab:null} quantifies the shrinking of the shadow. As $B$ runs from $0.1$ to $1.0$, the photon sphere contracts from $2.906M$ to $2.497M$ and the shadow radius from $5.068M$ to $4.486M$, a reduction of about eleven percent. This contraction arises from the same numerator that cools the horizon in Eq.~\eqref{TH_log_final}, since the logarithmic term deepens the potential well and pulls the unstable circular orbit inward. The photon-orbit angular velocity $\Omega_{\rm c}=\sqrt{f(r_{\rm ph})}/r_{\rm ph}$ rises with $B$, and so does the Lyapunov exponent $\lambda$, which measures how fast nearby null rays diverge from the photon sphere. A smaller shadow at fixed mass is the kind of signature that horizon-scale imaging targets, so the coupling range explored here maps onto a measurable shift in apparent size relative to a Schwarzschild hole of the same mass.

\begin{figure*}[tp]
  \centering
  \includegraphics[width=0.86\textwidth]{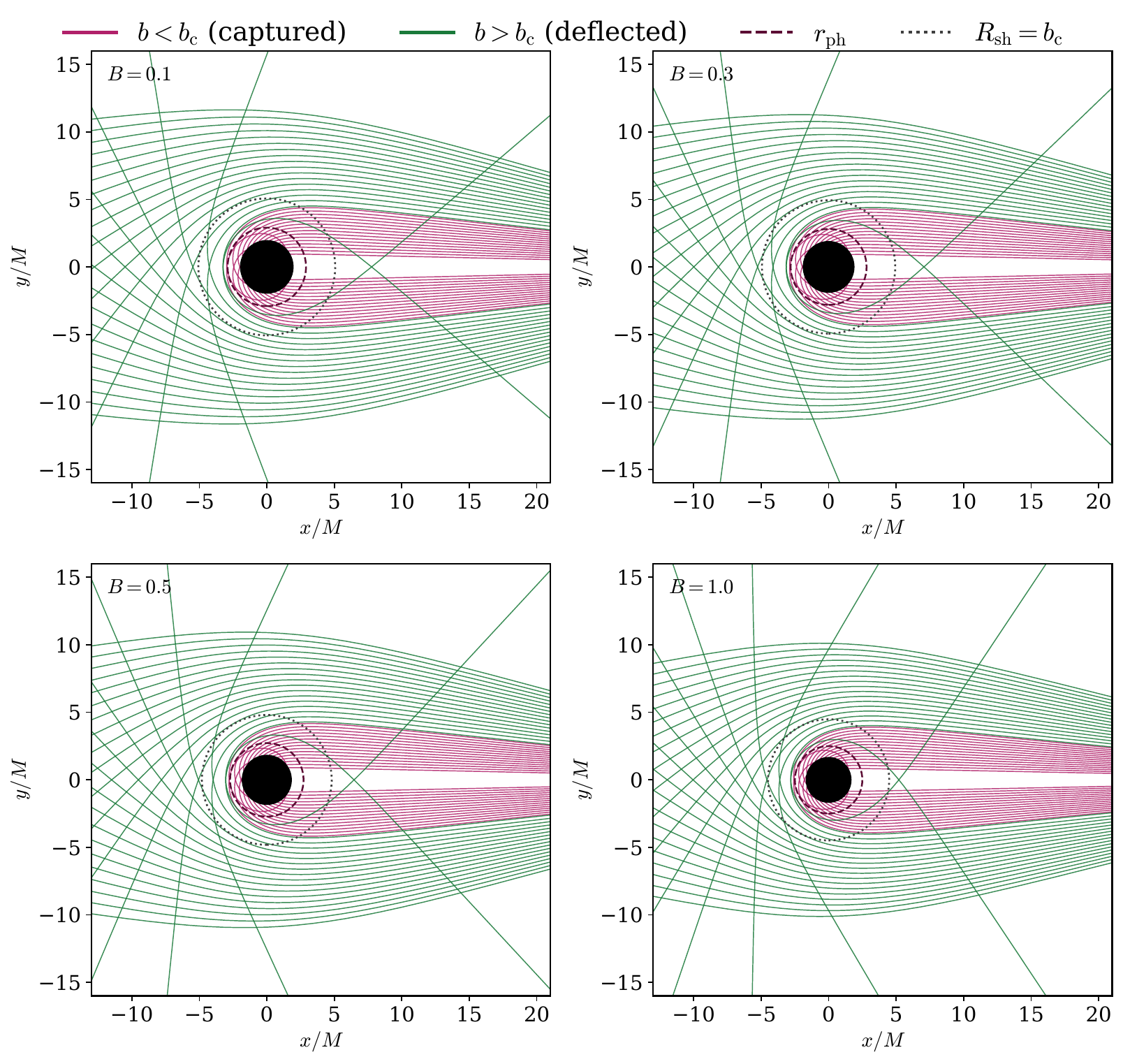}
  \caption{Equatorial null rays for $B=0.1,\,0.3,\,0.5,\,1.0$. Magenta rays with
    impact parameter $b<b_{\rm c}$ are captured; green rays with $b>b_{\rm c}$
    are deflected. The dashed circle marks the photon sphere $r_{\rm ph}$ and the
    dotted circle the shadow radius $R_{\rm sh}=b_{\rm c}$; the filled disk is the
    horizon. Rays are integrated from Eqs.~\eqref{eq:Vph}--\eqref{eq:bc}.}
  \label{fig:shadow}
\end{figure*}

The ray families in Fig.~\ref{fig:shadow} make the capture cross section explicit. Rays launched with $b<b_{\rm c}$ spiral through the photon sphere and fall in, while rays with $b>b_{\rm c}$ turn around and escape, and the dividing case grazes the unstable circular orbit. Comparing the four panels, the captured region tightens as $B$ grows, because the photon sphere and the shadow both move inward, which follows from the values in Table~\ref{tab:null}. The near-critical rays wind many times close to $r_{\rm ph}$, and this winding is controlled by $\lambda$, so the same quantity that sets the ringdown damping also sets the sharpness of the photon ring. The panels therefore give a geometric-optics counterpart to the eikonal correspondence developed in Sec.~\ref{isec9}.

\section{Timelike Geodesics, ISCO, and Zoom-Whirl Orbits}
\label{isec7}

Massive test particles probe the geometry closer to the horizon than photons do, and their orbits set the inner edge of an accretion disk. For a timelike geodesic in the equatorial plane, the radial equation is
\begin{equation}
\dot r^{2}=E^{2}-V_{\rm eff}(r),
\qquad
V_{\rm eff}(r)=f(r)\left(1+\frac{L^{2}}{r^{2}}\right),
\label{eq:Veff_time}
\end{equation}
where $E$ and $L$ are the specific energy and angular momentum. Circular orbits satisfy $V_{\rm eff}=E^{2}$ and $V_{\rm eff}'=0$, which give
\begin{equation}
E^{2}=\frac{f^{2}}{f-\tfrac12 r f'},
\qquad
L^{2}=\frac{r^{3}f'}{2f-r f'}.
\label{eq:EL_circ}
\end{equation}
The ISCO marks the transition from stable to unstable circular motion. It is fixed by the marginal-stability condition $dL^{2}/dr=0$, equivalently $V_{\rm eff}''=0$, which we solve numerically for each $B$. In the Schwarzschild limit the procedure returns $r_{\rm ISCO}=6M$, $E=\sqrt{8/9}\simeq0.9428$, which the code reproduces.

\begin{table*}[tp]
  \centering
  \setlength{\tabcolsep}{12pt}
  \renewcommand{\arraystretch}{1.5}
  \begin{tabularx}{\textwidth}{c c c c c c}
    \hline\hline
    \textbf{\boldmath $B$} & \textbf{\boldmath $r_{\rm ISCO}/M$} & \textbf{\boldmath $E_{\rm ISCO}$} & \textbf{\boldmath $L_{\rm ISCO}/M$} & \textbf{\boldmath $\eta\,(\%)$} & \textbf{\boldmath $\nu_{\varphi}\,M$} \\
    \hline
    0.1 & 5.7864 & 0.94092 & 3.38947 & 5.908 & 0.01129 \\
    0.3 & 5.5753 & 0.93897 & 3.30885 & 6.103 & 0.01176 \\
    0.5 & 5.3720 & 0.93696 & 3.22983 & 6.304 & 0.01226 \\
    0.7 & 5.1707 & 0.93484 & 3.15045 & 6.516 & 0.01279 \\
    1.0 & 4.8671 & 0.93135 & 3.02884 & 6.865 & 0.01366 \\
    \hline\hline
  \end{tabularx}
  \caption{ISCO of the logarithmic BH: radius
    $r_{\rm ISCO}$, specific energy $E_{\rm ISCO}$, specific angular momentum
    $L_{\rm ISCO}$, radiative efficiency $\eta=1-E_{\rm ISCO}$, and orbital
    frequency $\nu_{\varphi}$ at the ISCO. Geometric units, $M=1$, $Q=0.5$, $R_0=1$.}
  \label{tab:isco}
\end{table*}

Table~\ref{tab:isco} shows the inner disk edge moving inward with the coupling. The ISCO radius falls from $5.786M$ at $B=0.1$ to $4.867M$ at $B=1.0$, while the binding energy at the ISCO grows, so the radiative efficiency $\eta=1-E_{\rm ISCO}$ rises from about $5.9\%$ to $6.9\%$. This increase happens because the logarithmic term deepens the potential well of Eq.~\eqref{eq:Veff_time}, which lets a particle sit on a stable orbit closer to the horizon and release more binding energy before plunging. A more efficient inner disk at stronger coupling is a thermodynamically relevant statement, since $\eta$ sets the maximum fraction of rest mass that steady thin-disk accretion can radiate. The orbital frequency at the ISCO climbs in step, from $0.0113/M$ to $0.0137/M$, which feeds directly into the resonance analysis of Sec.~\ref{isec8}.

\begin{figure*}[tp]
  \centering
  \includegraphics[width=0.86\textwidth]{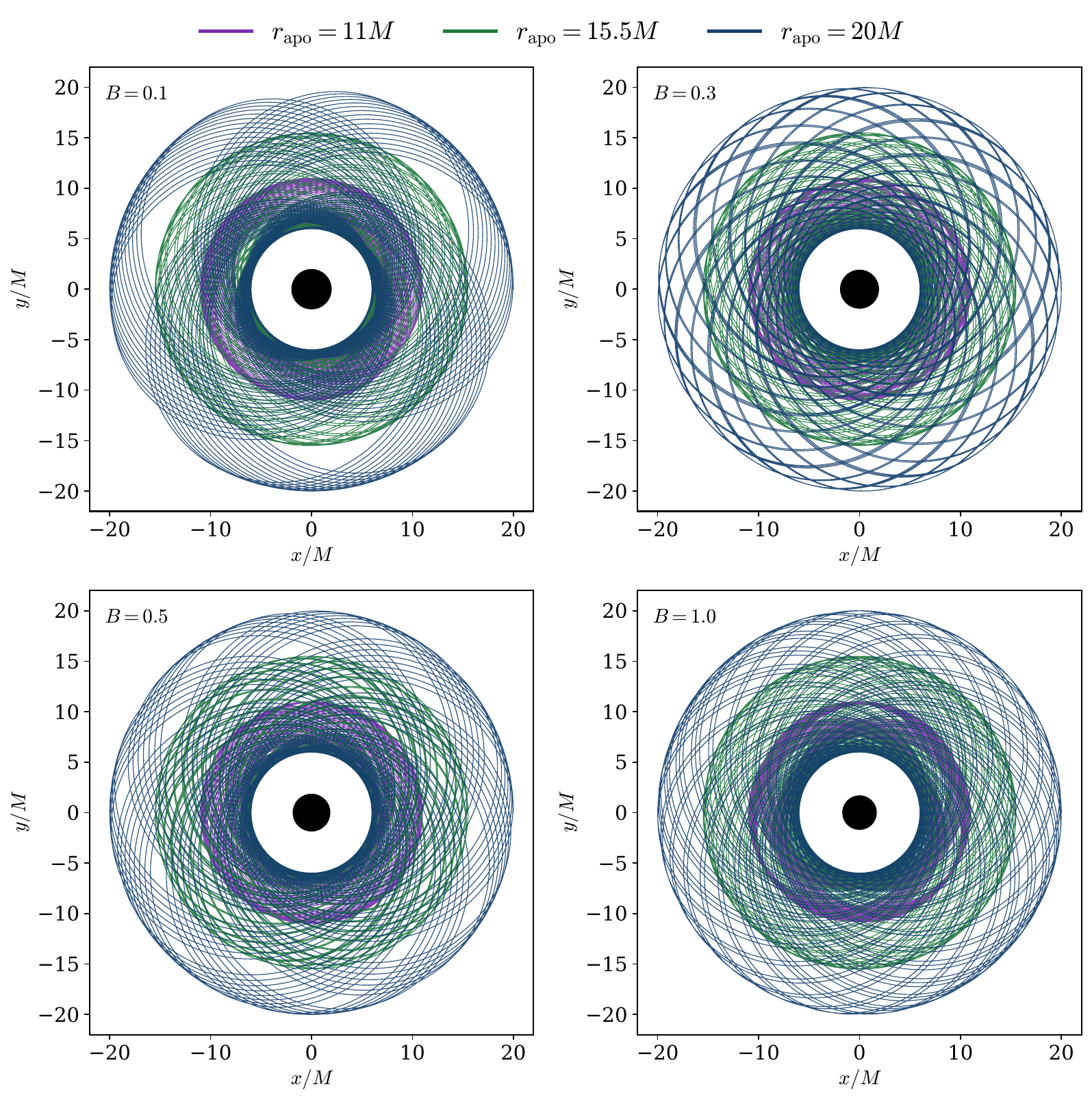}
  \caption{Bound timelike orbits for $B=0.1,\,0.3,\,0.5,\,1.0$ with fixed
    pericenter $r_{\rm per}=6M$ and apocenters $r_{\rm apo}=11M$ (purple),
    $15.5M$ (green), $20M$ (blue). Orbits are integrated from
    Eqs.~\eqref{eq:Veff_time} and~\eqref{eq:EL_circ}; the filled disk is the
    horizon.}
  \label{fig:zoomwhirl}
\end{figure*}

The bound orbits in Fig.~\ref{fig:zoomwhirl} display the zoom-whirl behaviour that strong-field precession produces. Each orbit alternates between a wide zoom near apocenter and a tight whirl near pericenter, and the accumulated periastron advance per revolution grows as $B$ increases. This growth arises because the deeper well shifts the ratio of the radial and azimuthal frequencies away from unity, so more azimuthal winding occurs between successive pericenter passages. Comparing panels, the innermost orbit whirls more times at $B=1.0$ than at $B=0.1$, which is the orbital-mechanics counterpart of the inward ISCO shift in Table~\ref{tab:isco}. Because the whirl count is sensitive to the near-horizon shape of $f$, extended tracking of a star or a hot spot on such an orbit would in principle separate the logarithmic geometry from a Reissner-Nordstr\"om one of the same mass and charge.

\section{Orbital Resonances, Eikonal Modes, and Characteristic Scales}
\label{isec8}

The circular-orbit structure of the exterior fixes a set of observables that a distant instrument can reach, and this section collects three of them. The orbital and epicyclic frequencies of nearly circular matter set the twin-peak resonance ratios seen in accreting systems \cite{Aliev:1980hz,Stella:1997tc,Abramowicz:2004je}. The same unstable photon orbit that bounds the shadow fixes the large-multipole ringdown through the eikonal correspondence \cite{Claudel:2000yi,Cardoso:2008bp}. The horizon, the photon sphere, the innermost stable orbit, and the shadow together organize the geometry into a small set of characteristic radii. We treat each in turn and show that the logarithmic coupling moves them in step rather than independently.

\subsection{Orbital Resonances and Quasiperiodic Oscillations}

The frequencies of nearly circular motion connect the geometry to a concrete X-ray observable. Accreting BHs and neutron stars show QPOs in their power spectra, and the high-frequency pairs often appear close to a $3\!:\!2$ ratio, which resonance models tie to the orbital and epicyclic frequencies of matter near the ISCO~\cite{stella1999,rogers2015frequency}. For the metric \eqref{eq:metric}, the orbital (azimuthal) frequency of a circular geodesic is
\begin{equation}
\nu_{\varphi}=\frac{1}{2\pi}\sqrt{\frac{f'(r)}{2r}},
\label{eq:nuphi}
\end{equation}
and, for a static spherically symmetric background, the vertical epicyclic frequency equals the orbital one, $\nu_{\theta}=\nu_{\varphi}$. The radial epicyclic frequency follows from the curvature of the effective potential \eqref{eq:Veff_time} about the circular orbit,
\begin{equation}
\nu_{r}=\frac{1}{2\pi}
\sqrt{\frac{f^{2}}{2E^{2}}\,
\left[-\frac{d^{2}}{dr^{2}}
\!\left(f+\frac{fL^{2}}{r^{2}}\right)\right]},
\label{eq:nur}
\end{equation}
with $E$ and $L$ evaluated on the orbit through Eq.~\eqref{eq:EL_circ}. We validated Eqs.~\eqref{eq:nuphi} and~\eqref{eq:nur} against the Schwarzschild relation $\nu_{r}=\nu_{\varphi}\sqrt{1-6M/r}$, which they reproduce, and which vanishes at $r=6M$ as it must.

\begin{table*}[tp]
  \centering
  \setlength{\tabcolsep}{12pt}
  \renewcommand{\arraystretch}{1.5}
  \begin{tabularx}{\textwidth}{c c c c c}
    \hline\hline
    \textbf{\boldmath $B$} & \textbf{\boldmath $\nu_{\varphi}\,M$} & \textbf{\boldmath $\nu_{r}\,M$} & \textbf{\boldmath $\nu_{\theta}\,M$} & \textbf{\boldmath $\nu_{\varphi}/\nu_{r}$} \\
    \hline
    0.1 & 0.00696 & 0.00365 & 0.00696 & 1.9069 \\
    0.3 & 0.00689 & 0.00378 & 0.00689 & 1.8249 \\
    0.5 & 0.00682 & 0.00388 & 0.00682 & 1.7555 \\
    0.7 & 0.00675 & 0.00398 & 0.00675 & 1.6944 \\
    1.0 & 0.00665 & 0.00412 & 0.00665 & 1.6141 \\
    \hline\hline
  \end{tabularx}
  \caption{Orbital ($\nu_{\varphi}$), radial-epicyclic ($\nu_{r}$), and
    vertical-epicyclic ($\nu_{\theta}$) frequencies at $r=8M$, and the
    twin-peak ratio $\nu_{\varphi}/\nu_{r}$, for $M=1$, $Q=0.5$, $R_0=1$.}
  \label{tab:qpo}
\end{table*}

Table~\ref{tab:qpo} evaluates the three frequencies at a fixed sample radius $r=8M$. At this radius the frequency ratio $\nu_{\varphi}/\nu_{r}$ decreases from $1.907$ to $1.614$ as $B$ grows, so the coupling drives the ratio toward and then just below the $3\!:\!2$ value. This shift occurs because the radial epicyclic frequency rises with $B$ while the orbital frequency drops slightly, and the two effects together compress the ratio. The physical consequence is that the radius at which a $3\!:\!2$ resonance sits moves inward with the coupling. We find it near $10.35M$ at $B=0.1$, near $9.64M$ at $B=0.5$, and near $8.79M$ at $B=1.0$. A twin-peak detection with an independent mass estimate would therefore constrain the logarithmic coupling, since the resonance radius and the associated frequency scale both respond to $B$.

To make the scale concrete we convert the dimensionless frequencies to physical units through $f=\hat\nu\,c^{3}/(2\pi GM)$, which for a stellar-mass hole gives $f=32.3\,\mathrm{kHz}\,(M_{\odot}/M)\,\hat\nu$ with $\hat\nu=\nu M$. Table~\ref{tab:qpo_hz} lists the pair evaluated at the $3\!:\!2$ resonance radius for a fiducial mass $M=10\,M_{\odot}$. The upper frequency runs from about $15$ to $18\,$Hz and the lower from about $10$ to $12\,$Hz as the coupling increases, and both scale inversely with the mass, so a heavier or lighter source shifts the whole pair rigidly. The two frequencies keep the $3\!:\!2$ ratio by construction, but their common scale rises with $B$ because the resonance radius moves inward toward the faster-orbiting region, which is the observable that a fit to a specific source would exploit.

\begin{table}[tbp]
  \centering
  \setlength{\tabcolsep}{8pt}
  \renewcommand{\arraystretch}{1.45}
  \begin{tabularx}{\columnwidth}{c c c c c}
    \hline\hline
    \textbf{\boldmath $B$} & \textbf{\boldmath $r_{3:2}/M$} & \textbf{\boldmath $f_{\varphi}$ [Hz]} & \textbf{\boldmath $f_{r}$ [Hz]} & \textbf{\boldmath $\Delta f$ [Hz]} \\
    \hline
    0.1 & 10.454 & 15.1 & 10.1 & 5.0 \\
    0.3 & 10.092 & 15.8 & 10.5 & 5.3 \\
    0.5 & \phantom{0}9.743 & 16.5 & 11.0 & 5.5 \\
    0.7 & \phantom{0}9.401 & 17.2 & 11.5 & 5.7 \\
    1.0 & \phantom{0}8.885 & 18.4 & 12.3 & 6.1 \\
    \hline\hline
  \end{tabularx}
  \caption{Twin-peak frequencies at the $3\!:\!2$ resonance radius $r_{3:2}$ for a
    fiducial mass $M=10\,M_{\odot}$, with $Q=0.5$, $R_0=1$. The frequencies scale
    as $1/M$ for other masses.}
  \label{tab:qpo_hz}
\end{table}

\subsection{Eikonal Quasinormal Modes from the Photon Sphere}
\label{isec9}

The field analysis of Sec.~\ref{isec4} and the geodesic analysis of Sec.~\ref{isec6} meet in the eikonal limit. For a static, spherically symmetric BH, the large-multipole QNM frequencies are set by the unstable circular null orbit: the real part by the orbital angular velocity at the photon sphere, and the imaginary part by the Lyapunov exponent that measures the instability of that orbit~\cite{cardoso2009}. The relation is
\begin{equation}
\omega_{\rm eik}
=\Omega_{\rm c}\,\ell
-i\left(n+\tfrac12\right)\lambda,
\label{eq:eikonal}
\end{equation}
with
\begin{equation}
\Omega_{\rm c}=\frac{\sqrt{f(r_{\rm ph})}}{r_{\rm ph}},
\qquad
\lambda=\sqrt{\frac{f(r_{\rm ph})\,
\bigl[2f(r_{\rm ph})-r_{\rm ph}^{2}f''(r_{\rm ph})\bigr]}
{2\,r_{\rm ph}^{2}}}.
\label{eq:OmLam}
\end{equation}
For the Dirac field the angular quantum number enters as $\ell=\kappa$. The correspondence is exact only in the geometric-optics regime, so we present it as a large-$\kappa$ estimate rather than a replacement for the WKB spectrum of Sec.~\ref{isec4}. Its value is that it ties the ringing frequency to the shadow, since both are built from the same photon sphere.

\begin{table*}[tp]
  \centering
  \setlength{\tabcolsep}{12pt}
  \renewcommand{\arraystretch}{1.5}
  \begin{tabularx}{\textwidth}{c c c c}
    \hline\hline
    \textbf{\boldmath $B$} & \textbf{\boldmath $\omega_{\rm eik}\,M\ (\ell{=}2)$} & \textbf{\boldmath $\omega_{\rm eik}\,M\ (\ell{=}3)$} & \textbf{\boldmath $\omega_{\rm eik}\,M\ (\ell{=}4)$} \\
    \hline
    0.1 & $0.3946-0.0973\,i$ & $0.5919-0.0973\,i$ & $0.7892-0.0973\,i$ \\
    0.3 & $0.4051-0.0988\,i$ & $0.6077-0.0988\,i$ & $0.8103-0.0988\,i$ \\
    0.5 & $0.4159-0.1004\,i$ & $0.6239-0.1004\,i$ & $0.8318-0.1004\,i$ \\
    0.7 & $0.4273-0.1020\,i$ & $0.6409-0.1020\,i$ & $0.8546-0.1020\,i$ \\
    1.0 & $0.4458-0.1045\,i$ & $0.6687-0.1045\,i$ & $0.8916-0.1045\,i$ \\
    \hline\hline
  \end{tabularx}
  \caption{Eikonal Dirac QNM frequencies from Eqs.~\eqref{eq:eikonal}
    and~\eqref{eq:OmLam}, for the fundamental tone $n=0$ at $\ell=2,3,4$, with
    $M=1$, $Q=0.5$, $R_0=1$.}
  \label{tab:eikonal}
\end{table*}

The correspondence also organizes the way the ringdown and the image respond to the coupling. Since $\Omega_{\rm c}$ and $\lambda$ are both built from the metric function and its curvature at $r_{\rm ph}$, a single inward shift of the photon sphere raises the real part of the frequency and the damping rate together, which is the pattern the table displays. The real part follows the shadow inversely, because a smaller photon orbit turns faster, while the damping follows the sharpness of the photon ring, because a more unstable orbit sheds nearby rays more quickly. This shared origin is what lets a measured ringdown and a measured shadow be checked for mutual consistency, and it is the reason we place the eikonal estimate next to the null-sector table rather than treating it as a separate calculation.

Table~\ref{tab:eikonal} lists the eikonal frequencies for the three lowest multipoles that enter the Dirac spectrum. The real part scales linearly with $\ell$, as Eq.~\eqref{eq:eikonal} requires, and it rises with $B$ at fixed $\ell$ because the photon-orbit angular velocity $\Omega_{\rm c}$ increases with the coupling, in step with the shadow contraction in Table~\ref{tab:null}. The imaginary part is common to all multipoles at fixed $n$, and it grows with $B$ from $0.0973/M$ to $0.1045/M$, tracking the Lyapunov exponent. Comparing with the low-$\kappa$ WKB values of Tables~\ref{tab:QNM_B01}--\ref{tab:QNM_B10}, the eikonal estimate captures the correct scale and the sign of the trends, while the exact fundamental modes differ because $\kappa=2,3,4$ are not yet in the strict geometric-optics regime. The message is that a measured ringdown frequency and a measured shadow size carry the same information about $B$, since both are set by the unstable photon orbit.

\subsection{Characteristic Scales and Their Ordering}
\label{isec9a}

Before turning to the remaining observables, we gather the length scales that the previous sections produced and examine how they move together. The horizon radius, the photon sphere, the ISCO, and the shadow radius are the four radii that organize the exterior geometry, and each one has been computed above from an independent condition. Collecting them for a common parameter set shows whether the logarithmic coupling rescales the whole geometry uniformly or distorts it.

\begin{table*}[tp]
  \centering
  \setlength{\tabcolsep}{5pt}
  \renewcommand{\arraystretch}{1.45}
  \begin{tabularx}{\textwidth}{c c c c c c c}
    \hline\hline
    \textbf{\boldmath $B$} & \textbf{\boldmath $r_h$} & \textbf{\boldmath $r_{\rm ph}$} & \textbf{\boldmath $r_{\rm ISCO}$} & \textbf{\boldmath $R_{\rm sh}$} & \textbf{\boldmath $\frac{r_{\rm ph}}{r_h}$} & \textbf{\boldmath $\frac{r_{\rm ISCO}}{r_h}$} \\
    \hline
    0.1 & 1.9310 & 2.9055 & 5.7864 & 5.0682 & 1.5047 & 2.9966 \\
    0.3 & 1.8651 & 2.8125 & 5.5753 & 4.9366 & 1.5080 & 2.9893 \\
    0.5 & 1.8017 & 2.7226 & 5.3720 & 4.8086 & 1.5111 & 2.9817 \\
    0.7 & 1.7387 & 2.6330 & 5.1707 & 4.6808 & 1.5143 & 2.9739 \\
    1.0 & 1.6435 & 2.4971 & 4.8671 & 4.4864 & 1.5193 & 2.9614 \\
    \hline\hline
  \end{tabularx}
  \caption{Characteristic radii in units of $M$: horizon $r_h$, photon sphere
    $r_{\rm ph}$, ISCO $r_{\rm ISCO}$, and shadow
    radius $R_{\rm sh}$, with the ratios $r_{\rm ph}/r_h$ and $r_{\rm ISCO}/r_h$.
    Here $Q=0.5$, $R_0=1$.}
  \label{tab:radii}
\end{table*}

\begin{figure}[tbp]
  \centering
  \includegraphics[width=\columnwidth]{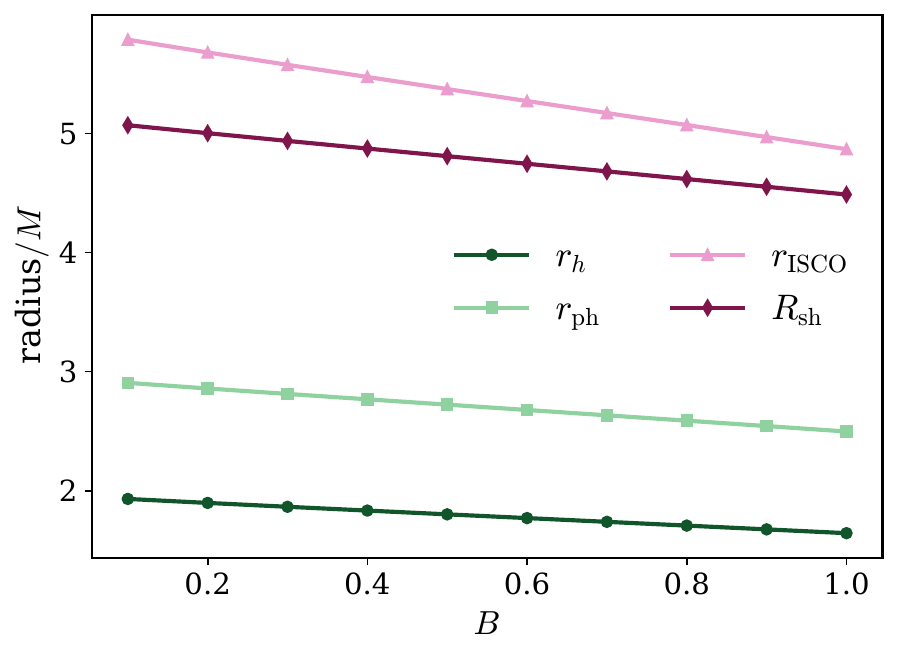}
  \caption{Characteristic radii $r_h$, $r_{\rm ph}$, $r_{\rm ISCO}$, and
    $R_{\rm sh}$ as functions of $B$, at $Q=0.5$, $R_0=1$. All four contract with
    the coupling while preserving their ordering.}
  \label{fig:radii}
\end{figure}

Table~\ref{tab:radii} and Fig.~\ref{fig:radii} show every radius shrinking as the coupling grows, and the strict ordering $r_h<r_{\rm ph}<R_{\rm sh}<r_{\rm ISCO}$ holds across the whole range. The contraction is not quite uniform, since the ratio $r_{\rm ph}/r_h$ rises slightly from $1.505$ to $1.519$ while $r_{\rm ISCO}/r_h$ falls from $2.997$ to $2.961$. This mild differential arises because the logarithmic term enters $f$ with a different radial weight than the mass term, so it pulls the near-horizon null orbit and the more distant timelike orbit by slightly different amounts. The upshot is that the strong coupling compresses the exterior geometry as a whole, and the small drift of the ratios is the fingerprint that separates the logarithmic background from a simple rescaling of a Reissner-Nordstr\"om hole. These are the scales that the resonance, ringdown, and deflection observables of this and the following section are built on.

\section{Weak and Strong Deflection of Light}
\label{isec9b}

The shadow of Sec.~\ref{isec6} is the boundary of a lensing map, and the deflection angle fills in that map for rays that escape. For a ray with impact parameter $b$ and closest approach $r_{\rm t}$, the bending angle in the equatorial plane is
\begin{equation}
\alpha(b)=2\int_{r_{\rm t}}^{\infty}
\frac{dr}{r^{2}\sqrt{\dfrac{1}{b^{2}}-\dfrac{f(r)}{r^{2}}}}-\pi,
\label{eq:deflection}
\end{equation}
where $r_{\rm t}$ is the largest root of $1/b^{2}=f(r_{\rm t})/r_{\rm t}^{2}$. The angle diverges as $b\to b_{\rm c}$, since the ray then approaches the photon sphere and winds indefinitely, and it decreases toward zero at large $b$. We evaluated Eq.~\eqref{eq:deflection} numerically and checked it against the Schwarzschild weak-field expansion $\alpha\simeq 4M/b+15\pi M^{2}/4b^{2}$, which it reproduces to better than $2\times10^{-3}$ at $b=30M$~\cite{Epstein:1980dw,Virbhadra2000,Bozza:2002zj,Keeton:2005jd}.

\begin{figure}[tbp]
  \centering
  \includegraphics[width=\columnwidth]{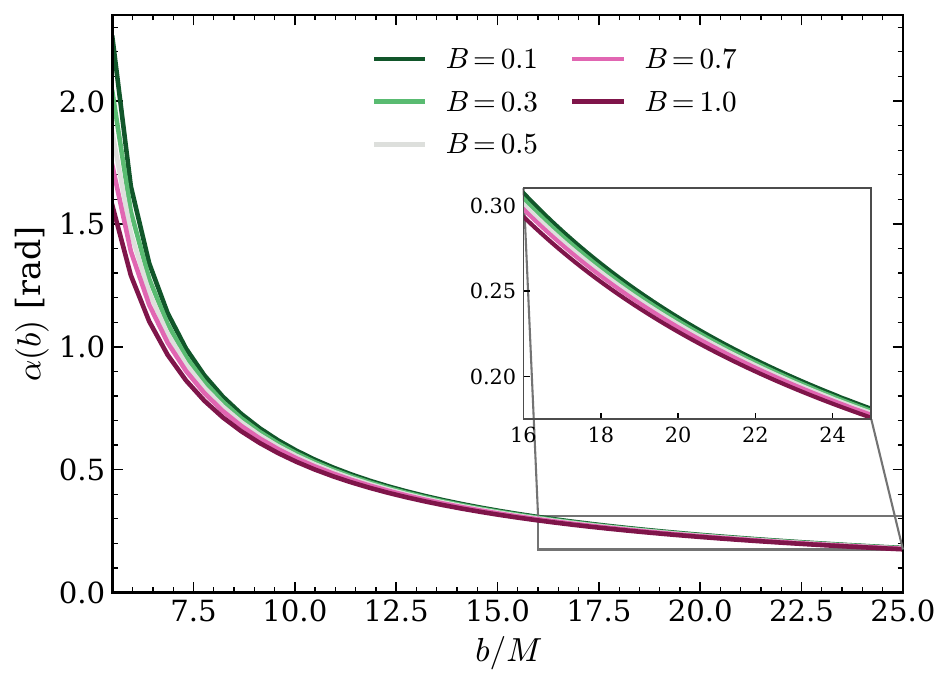}
  \caption{Deflection angle $\alpha(b)$ of Eq.~\eqref{eq:deflection} for
    $B=0.1,\,0.3,\,0.5,\,0.7,\,1.0$ (green to magenta), at $M=1$, $Q=0.5$,
    $R_0=1$. The angle grows sharply as $b$ approaches the critical value and
    falls off at large $b$.}
  \label{fig:deflection}
\end{figure}

Figure~\ref{fig:deflection} shows the bending angle rising steeply toward small impact parameter and flattening at large $b$. At fixed $b$ the angle decreases as $B$ grows, which may look surprising given that the strong coupling deepens the well, but it follows from the critical impact parameter contracting with $B$: a ray at a fixed $b$ then sits relatively farther from the (now smaller) photon sphere, so it bends less. The curves fan out most in the strong-deflection band near $b\sim6M$, where the logarithmic term acts most strongly, and they converge at large $b$ where the geometry is nearly Reissner-Nordstr\"om. This is the lensing counterpart of the shadow contraction in Table~\ref{tab:null}.

\begin{table*}[tp]
  \centering
  \setlength{\tabcolsep}{6pt}
  \renewcommand{\arraystretch}{1.45}
  \begin{tabularx}{\textwidth}{c c c c c c}
    \hline\hline
    \textbf{\boldmath $B$} & \textbf{\boldmath $b{=}6M$} & \textbf{\boldmath $b{=}8M$} & \textbf{\boldmath $b{=}10M$} & \textbf{\boldmath $b{=}15M$} & \textbf{\boldmath $b{=}20M$} \\
    \hline
    0.1 & 1.6129 & 0.8369 & 0.5804 & 0.3331 & 0.2345 \\
    0.3 & 1.5156 & 0.8137 & 0.5693 & 0.3293 & 0.2325 \\
    0.5 & 1.4338 & 0.7924 & 0.5589 & 0.3257 & 0.2307 \\
    0.7 & 1.3621 & 0.7724 & 0.5489 & 0.3222 & 0.2288 \\
    1.0 & 1.2685 & 0.7442 & 0.5346 & 0.3170 & 0.2261 \\
    \hline\hline
  \end{tabularx}
  \caption{Deflection angle $\alpha(b)$ in radians from Eq.~\eqref{eq:deflection},
    for $M=1$, $Q=0.5$, $R_0=1$. The angle falls with impact parameter and, at
    fixed $b$, decreases with the coupling $B$.}
  \label{tab:deflection}
\end{table*}

Table~\ref{tab:deflection} lists the angle for five impact parameters. In the strong band at $b=6M$ the bending is more than a radian and drops by about twenty percent as $B$ runs from $0.1$ to $1.0$, whereas at $b=20M$ the angle is near $0.23$ and barely moves. This spread means that the logarithmic coupling is most visible in the strong-lensing regime close to the photon sphere, where relativistic images form, and least visible in the weak-lensing tail. A resolved set of relativistic images, or a time-delay measurement between them, would therefore carry more information about $B$ than a single weak-lensing measurement at large separation.

\section{Hawking Emission and Barrow Thermodynamics of the Horizon}
\label{isec9c}

With the transmission of the fermionic field in hand, we turn to the thermal output of the horizon and to the thermodynamics that governs it. The section runs along two threads. The first follows the radiation itself, through the sparsity of the emitted quanta and the greybody-weighted power spectrum, and shows how the logarithmic coupling makes the hole fainter and more intermittent. The second recasts the horizon as a thermodynamic system carrying quantum-geometric corrections through the Barrow entropy, from which the temperature, the free energy, the effective pressure, and the heat capacity follow \cite{Ahmed:2025qgdef}. Read together, the two threads connect what leaves the hole to the thermal state of the horizon that emits it.

\subsection{Sparsity of the Hawking Flux}

The greybody analysis of Sec.~\ref{isec5} sets how much radiation escapes at each frequency. The sparsity sets how the escaping quanta are spaced in time. Hawking emission is not a steady stream: the average gap between successive quanta is large compared with the timescale of a single emitted wave, and this ratio is what makes the BH flux qualitatively different from that of a hot body of the same temperature~\cite{hawking1975particle}. A convenient dimensionless measure is
\begin{equation}
\eta=\frac{\lambda_{\rm th}^{2}}{A_{\rm eff}},
\qquad
\lambda_{\rm th}=\frac{1}{T_{H}},
\qquad
A_{\rm eff}=4\pi r_{h}^{2},
\label{eq:sparsity}
\end{equation}
where $\lambda_{\rm th}$ is the thermal wavelength scale and $A_{\rm eff}$ the horizon area. A larger $\eta$ means a sparser flux.

\begin{table}[tbp]
  \centering
  \setlength{\tabcolsep}{10pt}
  \renewcommand{\arraystretch}{1.45}
  \begin{tabularx}{\columnwidth}{c c c c}
    \hline\hline
    \textbf{\boldmath $B$} & \textbf{\boldmath $T_{H}\,M$} & \textbf{\boldmath $r_{h}/M$} & \textbf{\boldmath $\eta$} \\
    \hline
    0.1 & 0.04001 & 1.9310 & 13.33 \\
    0.3 & 0.04050 & 1.8651 & 13.95 \\
    0.5 & 0.04101 & 1.8017 & 14.58 \\
    0.7 & 0.04154 & 1.7387 & 15.25 \\
    1.0 & 0.04240 & 1.6435 & 16.39 \\
    \hline\hline
  \end{tabularx}
  \caption{Sparsity $\eta$ of Eq.~\eqref{eq:sparsity} with the Hawking
    temperature $T_{H}$ and horizon radius $r_{h}$, for $M=1$, $Q=0.5$, $R_0=1$.}
  \label{tab:sparsity}
\end{table}

Table~\ref{tab:sparsity} shows the flux growing sparser as the coupling increases, with $\eta$ rising from $13.3$ to $16.4$ over the range. This increase arises because the horizon shrinks faster than the temperature grows, so the ratio $\lambda_{\rm th}^{2}/A_{\rm eff}$ climbs with $B$. The value of $\eta$ stays well above unity throughout, which confirms that the emission remains sparse in the whole coupling range, in contrast to a laboratory black body where successive quanta overlap. Physically the strongly coupled hole radiates its (already reduced) power in more widely spaced quanta, so both the greybody suppression of Sec.~\ref{isec5} and the rising sparsity here point to a fainter and more intermittent signal at large $B$.

\subsection{Hawking Emission Power and Spectrum}
\label{isec9d}

The temperature, the greybody factor, and the sparsity combine into the quantity an observer would call the luminosity. For a fermionic field the number and energy fluxes are the greybody-weighted integrals over the Fermi-Dirac occupation \cite{Gray:2015pma,Visser:2015sal},
\begin{align}
\frac{dN}{dt}&=\frac{1}{2\pi}\sum_{\kappa}2\kappa
\int_{0}^{\infty}\frac{T_b(\omega,\kappa)}{e^{\omega/T_{H}}+1}\,d\omega,
\label{eq:Ndot}\\
P&=\frac{1}{2\pi}\sum_{\kappa}2\kappa
\int_{0}^{\infty}\frac{\omega\,T_b(\omega,\kappa)}{e^{\omega/T_{H}}+1}\,d\omega,
\label{eq:power}
\end{align}
where $2\kappa$ counts the multiplicity of the Dirac modes at fixed $\kappa$, and $T_b$ is the transmission bound of Eq.~\eqref{eq:greybody_bound}. We summed the three lowest multipoles $\kappa=1,2,3$, which carry the bulk of the flux at these temperatures, and integrated numerically.

\begin{figure}[ht!]
  \centering
  \includegraphics[width=\columnwidth]{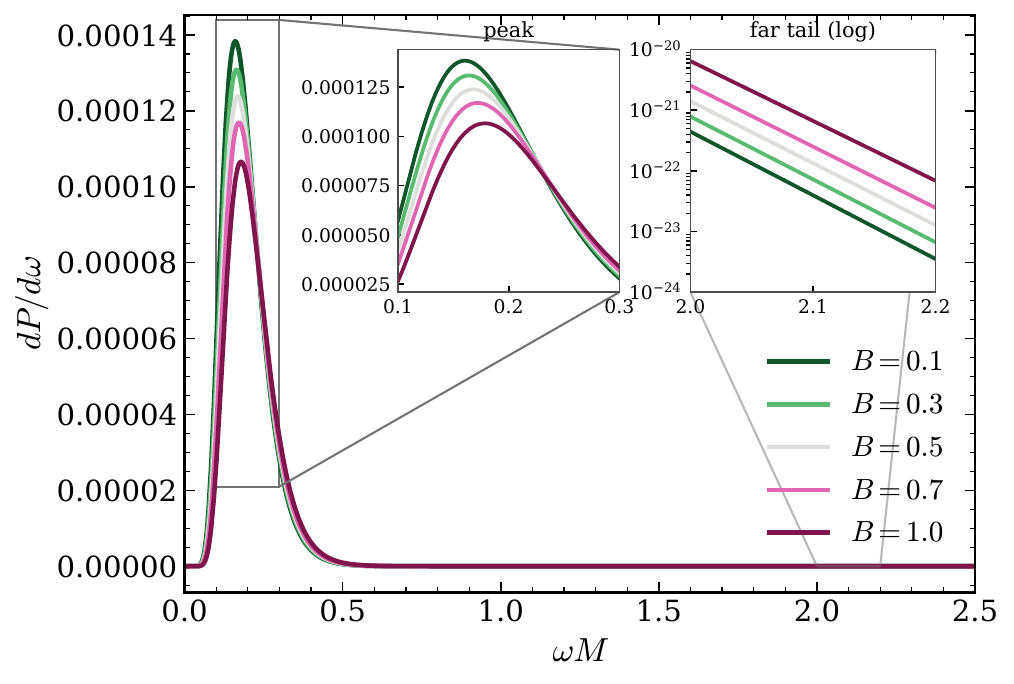}
  \caption{Power spectrum $dP/d\omega$ of the fermionic Hawking flux from the
    integrand of Eq.~\eqref{eq:power}, summed over $\kappa=1,2,3$, for
    $B=0.1,\,0.3,\,0.5,\,0.7,\,1.0$ (green to magenta), at $M=1$, $Q=0.5$,
    $R_0=1$.}
  \label{fig:spectrum}
\end{figure}

The spectra in Fig.~\ref{fig:spectrum} are single-humped, rising from the greybody-suppressed low-frequency side and falling on the thermal high-frequency side. As $B$ increases the peak lowers and shifts slightly to the right, because the greybody factor holds back more of the soft flux while the marginally higher temperature moves the thermal turnover outward. The area under each curve is the total power, so the visible loss of height with $B$ is the same statement as the falling luminosity in the table below. The spectrum thus encodes both the temperature through the position of the peak and the transmission through the shape of the low-frequency rise.

\begin{table}[tbp]
  \centering
  \setlength{\tabcolsep}{8pt}
  \renewcommand{\arraystretch}{1.45}
  \begin{tabularx}{\columnwidth}{c c c c}
    \hline\hline
    \textbf{\boldmath $B$} & \textbf{\boldmath $P\,M^{2}$} & \textbf{\boldmath $\dot{N}\,M$} & \textbf{\boldmath $\omega_{\rm peak}\,M$} \\
    \hline
    0.1 & $2.079\times10^{-5}$ & $1.224\times10^{-4}$ & 0.1611 \\
    0.3 & $1.999\times10^{-5}$ & $1.151\times10^{-4}$ & 0.1644 \\
    0.5 & $1.925\times10^{-5}$ & $1.084\times10^{-4}$ & 0.1678 \\
    0.7 & $1.851\times10^{-5}$ & $1.019\times10^{-4}$ & 0.1711 \\
    1.0 & $1.736\times10^{-5}$ & $9.213\times10^{-5}$ & 0.1778 \\
    \hline\hline
  \end{tabularx}
  \caption{Fermionic Hawking power $P$, particle emission rate $\dot{N}$, and
    spectral peak frequency $\omega_{\rm peak}$ from Eqs.~\eqref{eq:Ndot}
    and~\eqref{eq:power}, summed over $\kappa=1,2,3$, for $M=1$, $Q=0.5$,
    $R_0=1$.}
  \label{tab:power}
\end{table}

Table~\ref{tab:power} makes the luminosity trend quantitative. The emitted power falls by about sixteen percent as $B$ runs from $0.1$ to $1.0$, and the particle rate falls by a similar fraction, while the peak frequency climbs from $0.161/M$ to $0.178/M$. The drop in power arises because the greybody suppression outweighs the small rise in temperature, so fewer quanta escape per unit time even though each carries slightly more energy on average. This is the same conclusion reached separately from the transmission curves of Sec.~\ref{isec5} and the sparsity discussed above, and it gives a single number, the luminosity, that folds the temperature and the barrier together. A strongly coupled logarithmic hole is therefore both cooler in output and harder in spectrum than its weakly coupled counterpart.

\subsection{Thermodynamics via Barrow Entropy}
\label{isec10}

We now return to the horizon and study its thermal response with quantum-geometric corrections. The horizon radius $r_h$ is the largest positive root of $f(r_h)=0$, and for brevity we write $r_h\equiv r$ below. Solving the horizon condition for the mass parameter gives
\begin{equation}
M(r,Q,B)=\frac{r}{2}
+\frac{B Q^{2}}{2r}\ln\!\left(\frac{r}{r_{0}}\right)
+\frac{Q^{2}(1+5B)}{8r}.
\label{eq:mass_log}
\end{equation}
This reduces to the Reissner-Nordstr\"om mass relation once the logarithmic coupling is switched off. We verified Eq.~\eqref{eq:mass_log} by solving $f(r)=0$ symbolically for $M$.

To include quantum-geometric corrections at the horizon, we use the Barrow entropy
\begin{equation}
S_B=\left(\pi r^{2}\right)^{1+\frac{\Delta}{2}},
\qquad 0\leq \Delta \leq 1 ,
\label{eq:barrow_entropy_log}
\end{equation}
where $\Delta=0$ returns the Bekenstein-Hawking entropy and nonzero $\Delta$ parametrizes the fractal deformation of the horizon area~\cite{barrow2020,gomes2020thermodynamics}. Using the first law in the form $T_B=(\partial M/\partial S_B)_{Q,B,r_0}$, we obtain the Barrow-corrected temperature
\begin{equation}
T_B=
-\frac{\pi^{-1-\frac{\Delta}{2}}
\left(r^{2}\right)^{-\frac{\Delta}{2}}
\left[
4BQ^{2}\ln\!\left(\frac{r}{r_{0}}\right)
+BQ^{2}+Q^{2}-4r^{2}
\right]}
{8r^{3}(2+\Delta)} .
\label{eq:TB_log}
\end{equation}
The numerator shows that the logarithmic non-minimal coupling reaches the thermal sector both through the charge terms and through the scale-dependent piece $\ln(r/r_0)$. The temperature stays positive only where
\begin{equation}
4r^{2}>4BQ^{2}\ln\!\left(\frac{r}{r_{0}}\right)+(B+1)Q^{2}.
\end{equation}
We confirmed Eq.~\eqref{eq:TB_log} by symbolic differentiation of Eq.~\eqref{eq:mass_log} with respect to $S_B$ through the chain rule in $r$.

\begin{figure}[ht!]
  \centering
  \includegraphics[width=\columnwidth]{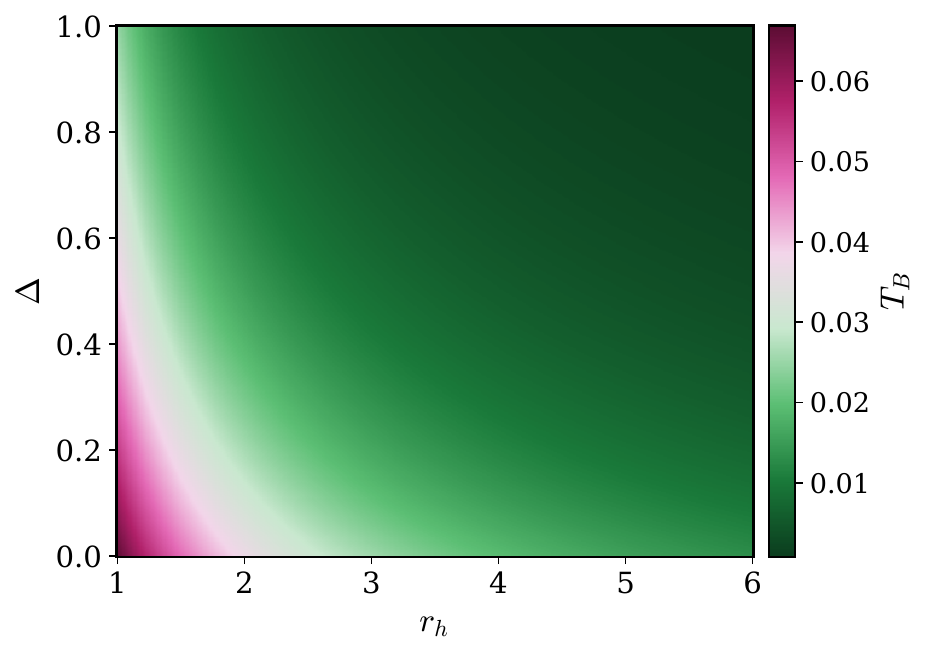}
  \caption{Barrow temperature $T_B$ of Eq.~\eqref{eq:TB_log} over the plane of
    horizon radius $r_h$ and Barrow parameter $\Delta$, at $Q=0.5$, $B=0.5$,
    $R_0=1$. The colour runs from low (green) to high (magenta).}
  \label{fig:T_B}
\end{figure}

The temperature map in Fig.~\ref{fig:T_B} falls off with horizon size and softens as $\Delta$ grows. Reading along $\Delta$ at fixed $r_h$, the deformation lowers $T_B$, because the prefactor $\pi^{-1-\Delta/2}(r^{2})^{-\Delta/2}$ in Eq.~\eqref{eq:TB_log} shrinks once $\pi r^{2}>1$, which holds for the horizons plotted. The fractal roughening therefore acts as a mild cooling of the horizon relative to the Bekenstein-Hawking case. The zero-temperature contour, where the numerator changes sign, separates the thermally active region from the forbidden one, and it moves outward as $B$ increases in line with the sign condition above.

The Helmholtz free energy is $\mathcal{F}_B=M-T_B S_B$. Substituting Eqs.~\eqref{eq:mass_log} and~\eqref{eq:TB_log} gives
\begin{widetext}
\begin{equation}
\mathcal{F}_B=
\frac{
4BQ^{2}(\Delta+3)\ln\!\left(\frac{r}{r_{0}}\right)
+\left[(1+5B)\Delta+11B+3\right]Q^{2}
+4r^{2}(\Delta+1)
}
{8r(2+\Delta)} .
\label{eq:FB_log}
\end{equation}
\end{widetext}
This quantity picks out the globally preferred branch, and the logarithmic term can shift the radius at which one branch overtakes another. Equation~\eqref{eq:FB_log} was checked symbolically against $M-T_B S_B$.

\begin{figure}[ht!]
  \centering
  \includegraphics[width=\columnwidth]{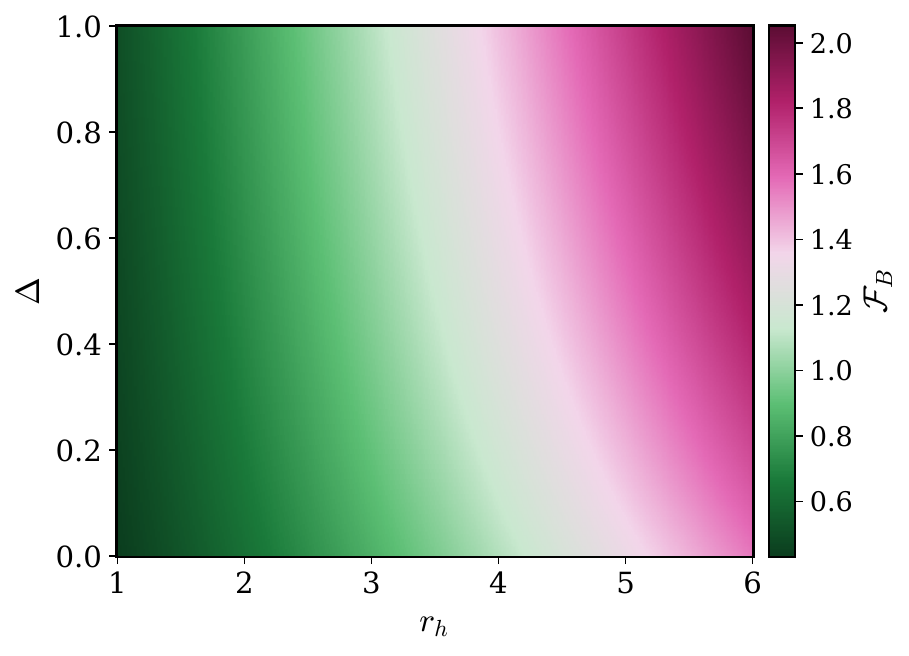}
  \caption{Barrow free energy $\mathcal{F}_B$ of Eq.~\eqref{eq:FB_log} over the
    $(r_h,\Delta)$ plane at $Q=0.5$, $B=0.5$, $R_0=1$ (green to magenta).}
  \label{fig:Free_FB}
\end{figure}

The free-energy surface of Fig.~\ref{fig:Free_FB} grows with the horizon radius and depends only weakly on $\Delta$ at large $r_h$. The mild $\Delta$ dependence arises because the deformation enters $\mathcal{F}_B$ through the ratios $(\Delta+3)/(2+\Delta)$ and $(\Delta+1)/(2+\Delta)$, which vary slowly over $0\le\Delta\le1$. The near-horizon rise is set instead by the charge and logarithmic terms in the numerator, so the globally preferred branch is selected mainly by $B$ and $Q$ rather than by the fractal parameter. This separation of roles guides the topological reading in Sec.~\ref{isec11}.

The effective pressure of the Barrow-deformed system follows from the equation of state as
\begin{widetext}
\begin{equation}
P_B=
\frac{
-4BQ^{2}(\Delta+3)\ln\!\left(\frac{r}{r_{0}}\right)
+\left[(-B-1)\Delta+B-3\right]Q^{2}
+4r^{2}(\Delta+1)
}
{32(2+\Delta)\pi r^{4}} .
\label{eq:PB_log}
\end{equation}
\end{widetext}
Here $P_B$ is an effective pressure read off from the logarithmic background, not a cosmological term. A symbolic comparison brings out a relation that we use below: the numerator of $P_B$ is identical to the numerator of the topological vector-field component $\phi_r$ of Sec.~\ref{isec11}, so that
\begin{equation}
P_B=\frac{\phi_r}{4\pi r^{2}} .
\label{eq:PB_phir}
\end{equation}
The pressure therefore vanishes exactly on the defect line where $\phi_r=0$.

\begin{figure}[ht!]
  \centering
  \includegraphics[width=\columnwidth]{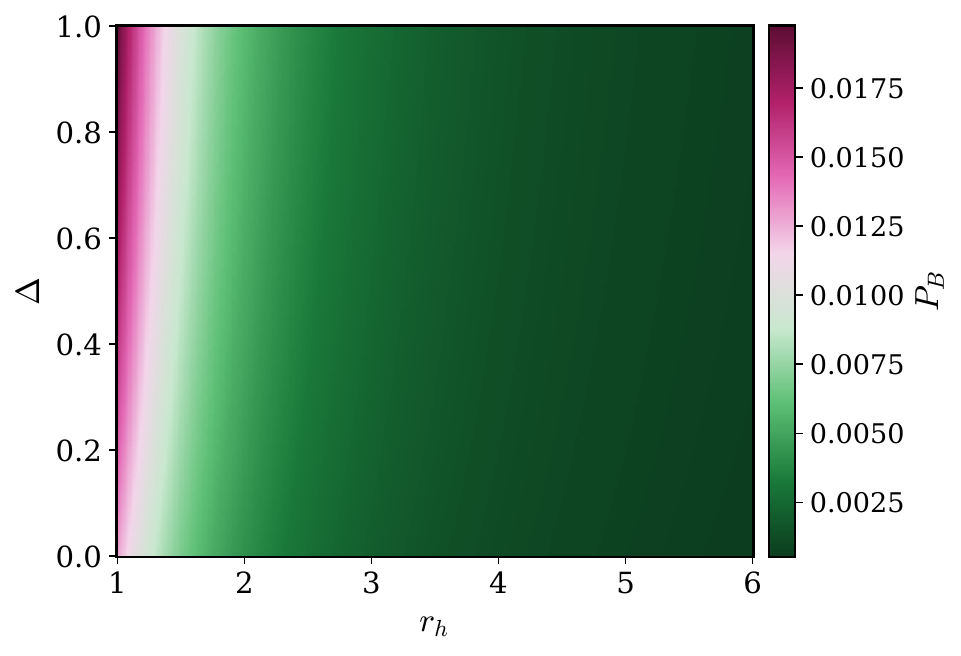}
  \caption{Effective pressure $P_B$ of Eq.~\eqref{eq:PB_log} over the
    $(r_h,\Delta)$ plane at $Q=0.5$, $B=0.5$, $R_0=1$ (green to magenta).}
  \label{fig:pressure_PB}
\end{figure}

The pressure map of Fig.~\ref{fig:pressure_PB} changes sign along a curve in the $(r_h,\Delta)$ plane, and that curve is where the identity~\eqref{eq:PB_phir} places the topological defects. Inside the sign-change locus the effective pressure is negative and outside it is positive, because the algebraic term $4r^{2}(\Delta+1)$ overtakes the charge and logarithmic terms only beyond the defect radius. This link is more than a coincidence of algebra. Since $P_B$ and $\phi_r$ share a numerator, the branch on which the pressure turns positive is the same branch that carries a positive winding number, so the sign of the effective pressure and the local thermodynamic stability are two readings of one condition. We return to this point after building the topological current.

Local stability is governed by the heat capacity
\begin{widetext}
\begin{equation}
C_B
=
-\frac{
4\left[
BQ^{2}\ln\!\left(\frac{r}{r_{0}}\right)
+\frac{(1+B)Q^{2}}{4}
-r^{2}
\right]
r^{2}
\left(r^{2}\right)^{\frac{\Delta}{2}}
\pi^{1+\frac{\Delta}{2}}
(2+\Delta)
}
{
4BQ^{2}(\Delta+3)\ln\!\left(\frac{r}{r_{0}}\right)
+\left[(\Delta-1)B+\Delta+3\right]Q^{2}
-4r^{2}(\Delta+1)
}.
\label{eq:CB_log}
\end{equation}
\end{widetext}
The zeros of the numerator give temperature-zero points, and the zeros of the denominator give second-order transition points, so the critical radii solve
\begin{equation}
4BQ^{2}(\Delta+3)\ln\!\left(\frac{r}{r_{0}}\right)
+\left[(\Delta-1)B+\Delta+3\right]Q^{2}
-4r^{2}(\Delta+1)=0 .
\label{eq:critical_CB_log}
\end{equation}
A branch with $C_B>0$ is locally stable and one with $C_B<0$ is locally unstable. Because of the logarithmic term, Eq.~\eqref{eq:critical_CB_log} does not reduce to a polynomial in $r$, so its roots are found numerically. We verified Eq.~\eqref{eq:CB_log} symbolically through $C_B=T_B(\partial S_B/\partial r)/(\partial T_B/\partial r)$.

\section{Topological and Geometric Characterization of the Phase Structure}
\label{isec11}

The heat capacity of the preceding section reports local stability one radius at a time. Here we read the global phase structure of the same horizon in two complementary geometric languages \cite{Weinhold:1975xej,ruppeiner1979thermodynamics,Sekhmani:2024vsu}. The first is the thermodynamic curvature of Ruppeiner geometry, whose sign reflects the effective interaction among the microscopic degrees of freedom \cite{Ruppeiner:1995zz,Ruppeiner:2010dzw,Wei:2015iwa}. The second is the topological method, in which the defects of a vector field built from the off-shell free energy mark the on-shell states, and their winding numbers separate stable from unstable branches. We then extend the construction to an exponential form of the Barrow entropy. The two readings agree with each other, and they tie back to the effective pressure through the identity established earlier.

\subsection{Ruppeiner Geometry}
\label{isec10b}

The heat capacity of Sec.~\ref{isec10} reports local stability. Thermodynamic geometry adds a complementary reading, in which the sign of a scalar curvature built from fluctuation theory reflects the effective interaction among the microscopic degrees of freedom~\cite{ruppeiner2007stability}. Working in the entropy representation with the Barrow entropy \eqref{eq:barrow_entropy_log} as the extensive variable, we form the curvature from the response function
\begin{equation}
\mathcal{R}\ \propto\ \frac{\partial^{2}T_{B}}{\partial S_B^{2}},
\label{eq:rupp_proxy}
\end{equation}
whose sign we track across the parameter plane. A positive sign is associated with a repulsive-type effective interaction and a negative sign with an attractive one, while a divergence signals a phase transition where the heat capacity also diverges.

\begin{table}[tbp]
  \centering
  \setlength{\tabcolsep}{8pt}
  \renewcommand{\arraystretch}{1.45}
  \begin{tabularx}{\columnwidth}{c c c c}
    \hline\hline
    \textbf{\boldmath $B$} & \textbf{\boldmath $\Delta=0$} & \textbf{\boldmath $\Delta=0.5$} & \textbf{\boldmath $\Delta=1$} \\
    \hline
    0.1 & $+2.34{\times}10^{-5}$ & $+1.99{\times}10^{-6}$ & $+1.58{\times}10^{-7}$ \\
    0.3 & $+2.27{\times}10^{-5}$ & $+1.96{\times}10^{-6}$ & $+1.56{\times}10^{-7}$ \\
    0.5 & $+2.17{\times}10^{-5}$ & $+1.89{\times}10^{-6}$ & $+1.51{\times}10^{-7}$ \\
    0.7 & $+2.09{\times}10^{-5}$ & $+1.83{\times}10^{-6}$ & $+1.47{\times}10^{-7}$ \\
    1.0 & $+2.00{\times}10^{-5}$ & $+1.78{\times}10^{-6}$ & $+1.44{\times}10^{-7}$ \\
    \hline\hline
  \end{tabularx}
  \caption{Thermodynamic-curvature proxy $\partial^{2}T_{B}/\partial S_B^{2}$ of
    Eq.~\eqref{eq:rupp_proxy} evaluated at $r=3M$, for $M=1$, $Q=0.5$, $R_0=1$.
    The sign is positive throughout, and the magnitude falls as the Barrow
    parameter $\Delta$ grows.}
  \label{tab:ruppeiner}
\end{table}

Table~\ref{tab:ruppeiner} shows the curvature proxy positive over the whole grid, so the effective microscopic interaction stays repulsive-type for the sampled radius, with no sign change and hence no fluctuation-driven transition there. The magnitude decreases sharply as $\Delta$ grows, from the $10^{-5}$ level at $\Delta=0$ to the $10^{-7}$ level at $\Delta=1$, because the Barrow entropy raises the power of the area in $S_B$ and thereby flattens the second derivative in Eq.~\eqref{eq:rupp_proxy}. The fractal deformation therefore pushes the system toward a nearly flat thermodynamic geometry, in which the microscopic interaction weakens even though its character is unchanged. This softening runs parallel to the mild cooling seen in the temperature map of Fig.~\ref{fig:T_B}, since both effects trace back to the same $\Delta$-dependent prefactor in $T_B$.

\subsection{Topological Vector Field and Winding Numbers}
\label{isec11top}

We now read the phase structure through the topological method~\cite{wei2020,wei2022,biro2018black}. The off-shell free energy is
\begin{equation}
\mathcal{F}=M-\frac{S_B}{\tau},
\label{eq:offshell_log}
\end{equation}
with $\tau$ an auxiliary inverse-temperature parameter, and the on-shell states are recovered at $\tau=T_B^{-1}$ \cite{wei2022,anand2025van}. Following the standard construction, we define the vector field \cite{Bai:2022klw}
\begin{equation}
\Phi=(\phi^r,\phi^\Theta),
\qquad
\phi^r=\frac{\partial \mathcal{F}}{\partial r},
\qquad
\phi^\Theta=-\cot\Theta\,\csc\Theta .
\label{eq:vector_log}
\end{equation}
Because $\phi^\Theta=0$ at $\Theta=\pi/2$, the defects sit on the equatorial line of the auxiliary space. For the logarithmic BH the radial component is
\begin{widetext}
\begin{multline}
\phi_r=
-\frac{3 B \,Q^{2} \ln\! \left(\frac{r}{r_{0}}\right)}{2 r^{2} \left(2+\Delta \right)}-\frac{B \Delta  Q^{2} \ln\! \left(\frac{r}{r_{0}}\right)}{2 r^{2} \left(2+\Delta \right)}-\frac{B \Delta  Q^{2}}{8 r^{2} \left(2+\Delta \right)}\\+\frac{B \,Q^{2}}{8 r^{2} \left(2+\Delta \right)}-\frac{\Delta  Q^{2}}{8 r^{2} \left(2+\Delta \right)}\\-\frac{3 Q^{2}}{8 r^{2} \left(2+\Delta \right)}+\frac{\Delta}{2 \left(2+\Delta \right)}+\frac{1}{2 \left(2+\Delta \right)} .
\label{eq:phir_log}
\end{multline}
\end{widetext}
The defects are therefore fixed by
\begin{widetext}
\begin{equation}
-4BQ^{2}(\Delta+3)\ln\!\left(\frac{r}{r_{0}}\right)
+\left[(-B-1)\Delta+B-3\right]Q^{2}
+4r^{2}(\Delta+1)=0 .
\label{eq:defect_log}
\end{equation}
\end{widetext}
This equation shows that the equilibrium structure is set by the interplay of the Barrow deformation $\Delta$, the charge $Q$, and the logarithmic coupling $B$. Compared with polynomial backgrounds, the logarithmic horizon can support a richer defect structure, because the algebraic $r^{2}$ term and the logarithmic term scale differently with $r$. We verified Eqs.~\eqref{eq:phir_log} and~\eqref{eq:defect_log} symbolically, and we confirmed that Eq.~\eqref{eq:defect_log} is the numerator identity behind $P_B=\phi_r/4\pi r^{2}$ of Sec.~\ref{isec10}.

The normalized field is $n^{a}=\phi^{a}/\sqrt{\phi^{b}\phi^{b}}$, and the topological current is
\begin{equation}
j^\mu=
\frac{1}{2\pi}
\epsilon^{\mu\nu\rho}\epsilon_{ab}
\partial_\nu n^a\partial_\rho n^b .
\end{equation}
The total charge is $W=\sum_{i}\omega_{i}$, where the winding number of an isolated defect on $\Theta=\pi/2$ is
\begin{equation}
\omega_i=
\mathrm{sign}
\left[
\left.
\frac{\partial \phi_r}{\partial r}
\right|_{r=r_i}
\right].
\label{eq:winding_log}
\end{equation}
A positive winding number marks a locally stable branch and a negative one an unstable branch.

The flow fields in Fig.~\ref{fig:topo_phi} locate the thermodynamic defects. Each panel shows one defect on the equatorial line $\Theta=\pi/2$, marked by the cyan disk, where $\phi_r$ changes sign. Moving from $B=0.1$ to $B=0.5$ at fixed $\Delta$, the defect shifts inward, because the logarithmic term in Eq.~\eqref{eq:defect_log} grows with $B$ and moves the root of the balance to smaller radius. The Barrow deformation, comparing the top and bottom rows, shifts the defect only slightly, which matches the weak $\Delta$ dependence already seen in the free energy of Fig.~\ref{fig:Free_FB}. Since the winding number reads the sign of $\partial\phi_r/\partial r$ at the defect, and since $P_B\propto\phi_r$, the single positive-winding defect in each panel is also the radius where the effective pressure turns positive, which ties the topological charge to the stability statement of Sec.~\ref{isec10}.

\subsection{Exponential Barrow Correction}
\label{isec11b}

We finally consider the exponential extension of the Barrow entropy. The off-shell free energy gains an exponential Barrow contribution, and the radial component of the vector field becomes
\begin{widetext}
\begin{multline}
\phi_r
=
-\frac{\left(r^{2}\right)^{-\Delta} \Delta  B \,Q^{2} \ln\! \left(\frac{r}{r_{0}}\right)}{2 r^{6} \left(2+\Delta \right) \pi^{2} \pi^{\Delta}}-\frac{B \Delta  Q^{2} \ln\! \left(\frac{r}{r_{0}}\right)}{2 r^{2} \left(2+\Delta \right)}\\-\frac{5 B \,Q^{2} \ln\! \left(\frac{r}{r_{0}}\right)}{4 r^{2} \left(2+\Delta \right)}-\frac{5 \left(r^{2}\right)^{-\Delta} Q^{2}}{16 r^{6} \left(2+\Delta \right) \pi^{2} \pi^{\Delta}}-\frac{5 Q^{2}}{16 r^{2} \left(2+\Delta \right)}-\frac{\left(r^{2}\right)^{-\Delta} \Delta  Q^{2} B}{8 r^{6} \left(2+\Delta \right) \pi^{2} \pi^{\Delta}}-\frac{B \Delta  Q^{2}}{8 r^{2} \left(2+\Delta \right)}-\frac{\left(r^{2}\right)^{-\Delta} \Delta  Q^{2}}{8 r^{6} \left(2+\Delta \right) \pi^{2} \pi^{\Delta}}\\-\frac{\Delta  Q^{2}}{8 r^{2} \left(2+\Delta \right)}-\frac{\left(r^{2}\right)^{-\Delta} Q^{2} B}{16 r^{6} \left(2+\Delta \right) \pi^{2} \pi^{\Delta}}-\frac{B \,Q^{2}}{16 r^{2} \left(2+\Delta \right)}+\frac{\left(r^{2}\right)^{-\Delta} \Delta}{2 r^{4} \left(2+\Delta \right) \pi^{2} \pi^{\Delta}}+\frac{\Delta}{2 \left(2+\Delta \right)}\\+\frac{3 \left(r^{2}\right)^{-\Delta}}{4 r^{4} \left(2+\Delta \right) \pi^{2} \pi^{\Delta}}+\frac{3}{4 \left(2+\Delta \right)}-\frac{5 \left(r^{2}\right)^{-\Delta} B \,Q^{2} \ln\! \left(\frac{r}{r_{0}}\right)}{4 r^{6} \left(2+\Delta \right) \pi^{2} \pi^{\Delta}}.
\label{eq:phir_exp_log}
\end{multline}
\end{widetext}
For $r>0$ the factor $\pi^{-2-\Delta}(r^{2})^{-\Delta}+r^{4}$ is strictly positive, so it adds no roots. The physical defects are then set by
\begin{widetext}
\begin{equation}
\left(\Delta+\frac{5}{2}\right)BQ^{2}\ln\!\left(\frac{r}{r_{0}}\right)
+\frac{\left(\frac{5}{2}+(1+B)\Delta+\frac{B}{2}\right)Q^{2}}{4}
-r^{2}\left(\Delta+\frac{3}{2}\right)=0 .
\label{eq:defect_exp_log}
\end{equation}
\end{widetext}
\begin{figure*}[ht!]
  \centering
  \includegraphics[width=0.92\textwidth]{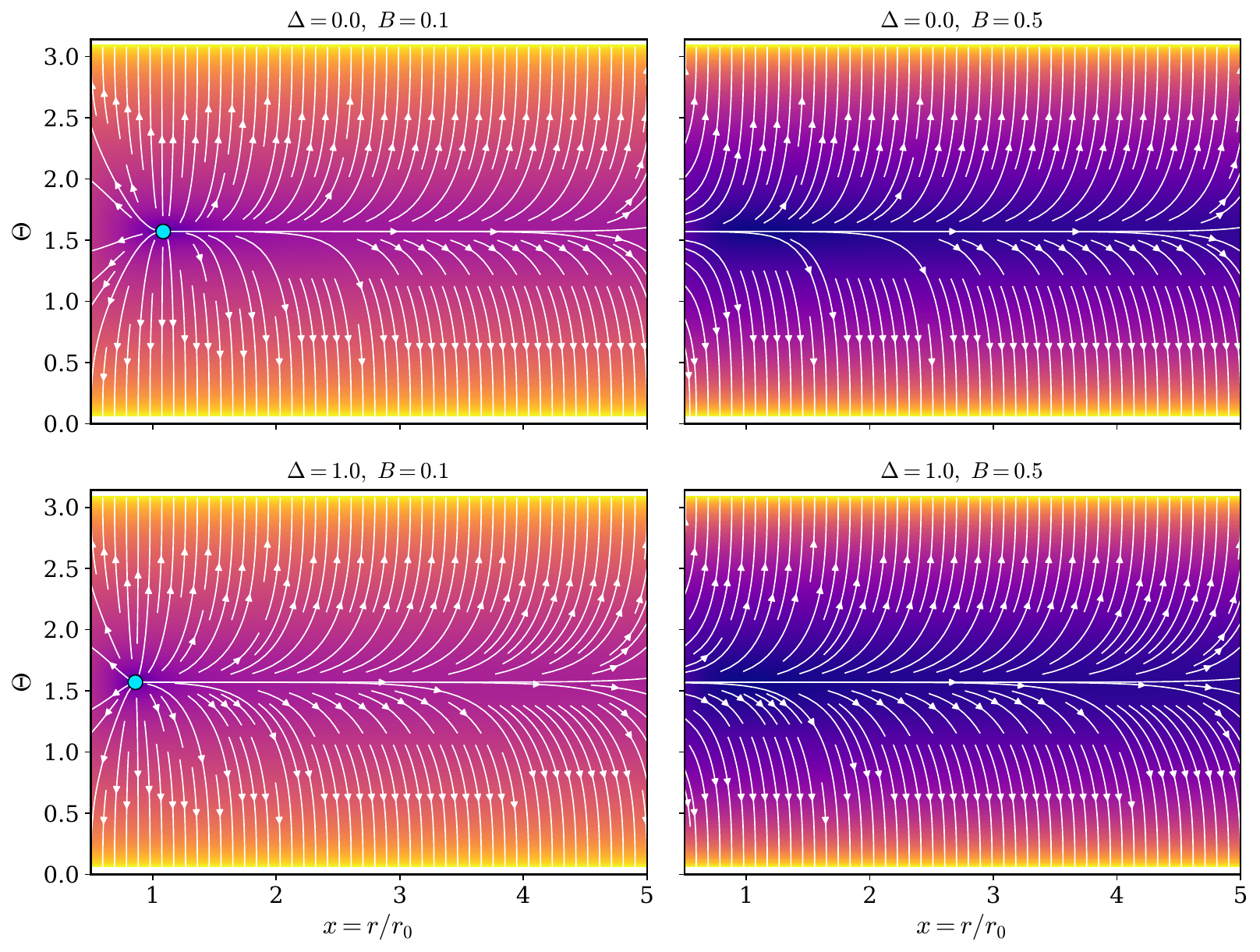}
  \caption{Normalized vector field $n=(n^{r},n^{\Theta})$ of
    Eq.~\eqref{eq:vector_log} in the $(x=r/r_0,\Theta)$ plane for
    $(\Delta,B)=(0,0.1),(0,0.5),(1,0.1),(1,0.5)$, at $Q=0.5$, $R_0=1$. Cyan disks
    mark the defects on $\Theta=\pi/2$ where $\phi_r=0$. Colour encodes the field
    magnitude on a logarithmic plasma colour scale (dark low to bright high), with white streamlines.}
  \label{fig:topo_phi}
\end{figure*}
The exponential correction changes the location and possible number of defects through the coefficients that multiply the logarithmic and algebraic terms. Since Eq.~\eqref{eq:defect_exp_log} is transcendental, its roots are found numerically for fixed $(B,Q,r_0,\Delta)$, and the sign of $\partial\phi_r/\partial r$ at each root then fixes the winding number and the local stability. We checked the strictly positive factor and the reduced defect condition~\eqref{eq:defect_exp_log} symbolically.

\begin{figure*}[ht!]
  \centering
  \includegraphics[width=0.92\textwidth]{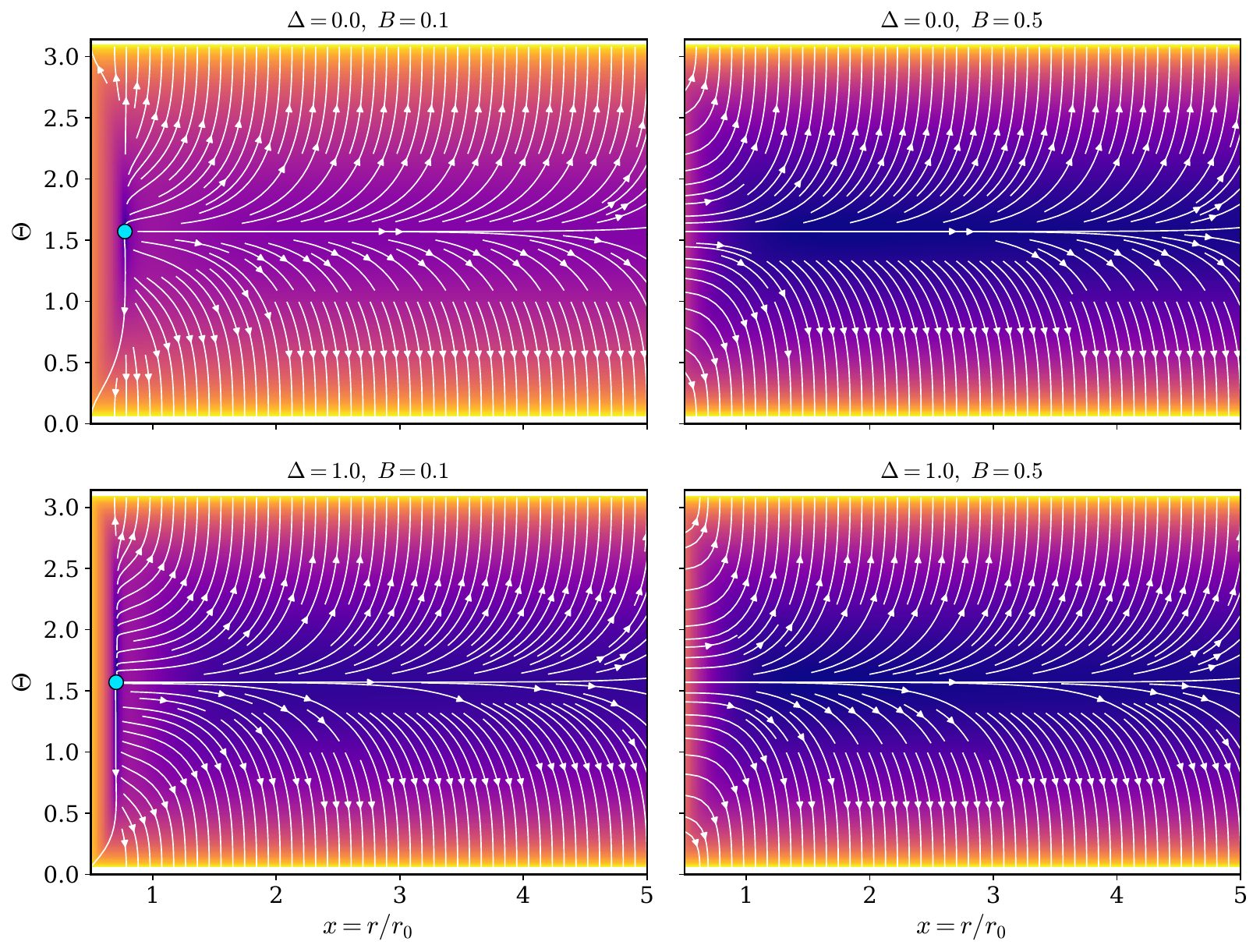}
  \caption{Normalized vector field for the exponential Barrow correction,
    Eq.~\eqref{eq:phir_exp_log}, in the $(x=r/r_0,\Theta)$ plane for the same
    parameter sets as Fig.~\ref{fig:topo_phi}. Colour encodes the field magnitude
    on a logarithmic plasma colour scale (dark low to bright high), with white streamlines.}
  \label{fig:topo_exp}
\end{figure*}

The exponential-correction fields in Fig.~\ref{fig:topo_exp} keep the equatorial defect structure but redistribute the flow lines relative to Fig.~\ref{fig:topo_phi}. The near-axis region is where the exponential terms $(r^{2})^{-\Delta}/r^{6}$ act most strongly, because those terms decay quickly with $r$ and dominate only at small radius. Away from the axis the field reverts to the pattern of the plain Barrow case, since the exponential contribution is then subdominant. The upshot is that the exponential deformation reshapes the inner flow and can move the defect radius, while the count of equatorial defects in the plotted window is unchanged, so the qualitative stability picture set by the winding number survives the extension \cite{wei2020,Fang:2022rsb,Fan:2022bsq}.

\section{Discussion and Observational Reach}
\label{isec11a}

It is useful to step back and ask why the many observables of this paper move together the way they do. The common thread is the near-horizon slope of the metric function. Raising the coupling $B$ at fixed charge and mass steepens $f'$ close to the horizon through the logarithmic term, and almost every quantity we computed is built from that slope or from the potential it generates. The Hawking temperature is $f'(r_{+})/4\pi$, so it climbs with $B$. The surface gravity that sets the temperature is the same slope that deepens the null and timelike potential wells, which pulls the photon sphere, the shadow, and the innermost stable orbit inward. The barrier that governs the Dirac ringing and the greybody transmission rises with the same steepening, so the fundamental mode lives longer and the soft radiation is held back. Seen this way, the cooler thermal output, the smaller shadow, the more efficient inner disk, the sparser flux, and the longer-lived ringdown are not five independent facts but five readings of one change in the near-horizon geometry.

A second thread runs through the thermodynamic and topological sectors. The Barrow deformation enters the temperature, the free energy, and the curvature proxy through a single $\Delta$-dependent prefactor, which is why it cools the horizon, flattens the thermodynamic geometry, and shifts the defect only mildly, all at once, while leaving the branch selection to the charge and the coupling. The identity $P_B=\phi_r/4\pi r^{2}$ then ties the effective pressure to the topological defect line, so that the sign of the pressure and the local stability are one statement. This is the tightest link in the paper between the exterior thermodynamics and the phase topology, and it is the relation we most want to understand from first principles.

The limitations should be stated plainly. The QNM spectrum rests on the third-order WKB method, which is reliable for the low overtones of a single-barrier potential but not for the highly damped modes or the near-extremal regime, so the frequencies we quote are the astrophysically relevant ones rather than the full spectrum. The greybody results are strict lower bounds rather than exact transmission coefficients, and the emitted power that follows is therefore a bound-based estimate. The eikonal correspondence is a large-multipole statement, and we present it as such. The background is static and spherically symmetric, so the shadow and the orbital frequencies lack the spin dependence that a real accreting system shows. None of these caveats changes the direction of the trends, but each marks a place where a sharper tool would sharpen the numbers.

The observables above map onto quantities that current and near-term facilities measure, and it is worth stating the reach in rough terms. The shadow radius contracts from $5.07M$ to $4.49M$ as $B$ runs from $0.1$ to $1.0$, a fractional change of about eleven percent. Horizon-scale imaging of the two resolved sources constrains the shadow size at roughly the ten percent level once the mass and distance are supplied independently, so the upper part of this coupling range sits near the current sensitivity, and tighter future imaging would narrow it further~\cite{EHT2019,vagnozzi2023}. A measured shadow that agrees with the Schwarzschild value at the few percent level would already disfavour the strong-coupling end.

The twin-peak resonance of Sec.~\ref{isec8} offers an independent handle. The $3\!:\!2$ commensurability that several microquasars display near their high-frequency QPOs fixes a resonance radius, and we find that radius moving inward from about $10.3M$ to $8.8M$ across the coupling range, with the associated frequency scale rising as the ISCO contracts~\cite{stella1999,rogers2015frequency}. For a source with an independent dynamical mass, the observed pair of frequencies then translates into a band in the coupling, since both the ratio and the absolute scale respond to $B$. The ringdown gives a third route through the eikonal frequencies of Sec.~\ref{isec9}, which share the photon-sphere origin of the shadow, so a joint fit of the imaged size and a measured damped frequency would be consistent only for a common value of $B$. These three probes are not equally mature, but they are independent, and their agreement or tension is what would eventually pin the logarithmic coupling rather than merely bound it.

\section{Conclusion}
\label{isec12}

We have followed the charged BH of Einstein-Maxwell theory with a non-minimal $\ln(R)F^{2}$ coupling from its horizon out to the orbits and radiation that a distant observer could measure. The Hamilton-Jacobi form of the fermionic tunnelling method fixed the temperature straight from the near-horizon metric, and the logarithmic coupling entered it through both the parameter $B$ and the scale $r_0$, cooling the horizon relative to the charged case at the same size. Treating a massless Dirac field as a probe, we computed the QNM spectrum with the third-order WKB method and read off a quality factor whose balance between ringing and decay depends on $B$, $\kappa$, and the overtone number together rather than on any one of them.

The exterior physics filled in the rest of the picture. We bounded the Dirac greybody factor and found that stronger coupling makes the hole a more selective emitter, holding back the soft part of the fermionic spectrum while leaving the hard part nearly untouched. The null geodesics gave a photon sphere and a shadow that both contract as $B$ grows, by about eleven percent over the range studied, and the timelike geodesics gave an ISCO that moves inward while its radiative efficiency climbs from roughly six to seven percent. The bound orbits showed zoom-whirl precession that sharpens with the coupling, and the orbital and epicyclic frequencies pushed the twin-peak resonance ratio toward and just below $3\!:\!2$, with the resonance radius moving inward as $B$ increases. Reading the eikonal QNMs off the same photon sphere tied the ringing frequency to the shadow, since both are built from the one unstable circular null orbit.

On the thermodynamic side, the Barrow entropy carried the quantum-geometric correction into the temperature, the free energy, the effective pressure, and the heat capacity. The fractal deformation cooled the horizon and shifted the stable branch only mildly, while the charge and the logarithmic coupling did most of the work of selecting the preferred state. The topological reading of the off-shell free energy located the defects on the equatorial line and assigned them winding numbers, and it showed the equilibrium set by the competition among the Barrow parameter, the charge, and the logarithmic coupling. One result ties the two halves of the paper together: the effective pressure and the topological vector field share a numerator, so $P_B=\phi_r/4\pi r^{2}$, and the pressure vanishes exactly on the defect line. The sign of the effective pressure and the local stability are then two readings of one condition. The exponential extension of the Barrow entropy reshaped the inner flow and could move the defect radius while leaving the equatorial defect count unchanged in the window we plotted.

Taken together, the results give a coherent account of how a single non-minimal coupling propagates through the physics of a charged BH. The logarithmic term leaves a scale in the metric, and that scale sets the near-horizon slope of the lapse, which in turn fixes the temperature, the barrier for perturbations, the photon sphere, and the circular-orbit structure. Because these quantities share one source, the coupling dependence is correlated across the thermal, radiative, and orbital sectors, and a measurement in one sector predicts the direction of the shift in the others. On the thermodynamic side, the Barrow deformation and the topological reading show the same economy, with a single prefactor controlling the cooling and the flattening of the thermodynamic geometry, and with the effective pressure and the defect line joined by one identity. The paper therefore does more than catalogue observables for a new background: it shows that they are tied together by the geometry, and it verifies each link both numerically against known limits and symbolically against the closed forms.

Many promising lines of research arise directly from these results, and we highlight them as specific next steps rather than as broad, open-ended possibilities. The eikonal correspondence of Sec.~\ref{isec9} invites a full time-domain integration of the Dirac field, which would test how far the geometric-optics estimate tracks the exact ringdown at low multipoles and would give a waveform to compare against the projected sensitivity of next-generation detectors in the relevant frequency band. The shadow contraction of Sec.~\ref{isec6} can be turned into a quantitative bound by folding the measured angular size of the imaged sources, together with independent mass and distance estimates, into a fit for $B$, which would place the logarithmic coupling on the same footing as other horizon-scale constraints. The twin-peak ratios of Sec.~\ref{isec8} similarly become a constraint once a resonance model is fixed and an independent mass is supplied, since the resonance radius and the frequency scale both respond to the coupling. On the thermodynamic side, the identity $P_B=\phi_r/4\pi r^{2}$ deserves a derivation from first principles that clarifies whether it is special to the logarithmic background or a feature of Barrow-deformed charged holes more broadly, and the extended phase space with the exponential correction can be mapped for swallowtail structure and first-order transitions. Finally, the rotating generalization of this geometry would let the same set of observables, the shadow, the ISCO, the quasiperiodic frequencies, and the ringdown, be computed with spin, which is the form in which they are actually measured~\cite{al-Badawi:2024pdx}. We intend to pursue these tasks in future work.

\begin{acknowledgments}
\.{I}.S. and E.S. thank the Eastern Mediterranean University, T\"UB\.{I}TAK, ANKOS, and SCOAP3 for academic support, and acknowledge the networking support of COST Actions CA22113 (``Fundamental challenges in theoretical physics''), CA21106 (``COSMIC WISPers in the Dark Universe''), and CA23130 (``Bridging high and low energies in search of quantum gravity'').
\end{acknowledgments}

\section*{Data Availability Statement}
The analytical expressions required to reproduce every result are given explicitly in the main text. The computational worksheets used to verify the closed-form relations and to generate the figures and tables are available from the corresponding author upon reasonable request.

\bibliographystyle{apsrev4-2}
\bibliography{finalref}

\end{document}